\documentclass[twocolumn,times]{aastex631}

\usepackage[OT2, T1]{fontenc}
\usepackage{amsmath}

\def\pp{\raise.22ex\hbox{{\footnotesize +}}\raise.22ex\hbox{\footnotesize +}}

\graphicspath{{./}{fig/}}
\usepackage{CJK}         

\shorttitle{3d-rmhd simulations for sub- and near-critical accretion}
\shortauthors{Huang et al.}

\begin{document}
\begin{CJK*}{UTF8}{gbsn}

\title{Global Three-Dimensional Radiation Magnetohydrodynamic Simulations of Accretion onto a Stellar Mass Black Hole at Sub- and Near-critical Accretion Rates}

\author[0000-0001-8674-2336]{Jiahui Huang}
\affiliation{Department of Engineering Physics, Tsinghua University, Beijing 100084, China}

\author[0000-0002-2624-3399]{Yan-Fei Jiang (姜燕飞)}
\email{yjiang@flatironinstitute.org}
\affiliation{Center for Computational Astrophysics, Flatiron Institute, New York, NY 10010, USA}

\author[0000-0001-7584-6236]{Hua Feng}
\email{hfeng@tsinghua.edu.cn}
\affiliation{Department of Astronomy, Tsinghua University, Beijing 100084, China}
\affiliation{Department of Engineering Physics, Tsinghua University, Beijing 100084, China}

\author{Shane W. Davis}
\affiliation{Department of Astronomy, University of Virginia, Charlottesville, VA 22904, USA}

\author{James M. Stone}
\affiliation{School of Natural Sciences, Institute for Advanced Study, Princeton, NJ
08544, USA}

\author{Matthew J. Middleton}
\affiliation{Department of Physics \& Astronomy, University of Southampton, Southampton, SO17 1BJ, UK}

\begin{abstract} 
We present global 3D radiation magnetohydrodynamical simulations of accretion onto a 6.62 solar mass black hole with quasi-steady state accretion rates reaching 0.016 to 0.9 times the critical accretion rate, which is defined as the accretion rate to power the Eddington luminosity assuming a 10\% radiative efficiency, in different runs. The simulations show no sign of thermal instability over hundreds of thermal timescales at 10~$r_{\rm g}$. The energy dissipation happens close to the mid-plane in the near-critical runs and near the disk surface in the low accretion rate run. The total radiative luminosity inside $\sim$20~$r_{\rm g}$ is about 1\% to 30\% the Eddington limit, with a radiative efficiency of about 6\% and 3\%, respectively, in the sub- and near-critical accretion regimes. In both cases, self-consistent turbulence generated by the magnetorotational instability (MRI) leads to angular momentum transfer, and the disk is supported by magnetic pressure. Outflows from the central low-density funnel with a terminal velocity of $\sim$0.1$c$ are seen only in the near-critical runs. We conclude that these magnetic pressure dominated disks are thermally stable and thicker than the $\alpha$ disk, and the effective temperature profiles are much flatter than that in the $\alpha$ disks. The magnetic pressure of these disks are comparable within an order of magnitude with the previous analytical magnetic pressure dominated disk model.
\end{abstract}

\section{Introduction}
\label{sec:intro}
 
X-ray binaries are among the most luminous X-ray objects in the Milky Way and nonactive galaxies \citep{Remillard2006}.  Their total luminosity is scaled with the star formation rate and total stellar mass of the host galaxy \citep{Gilfanov2004,Mineo2012}.  They are responsible for the heating of the intergalactic medium during the epoch of reionization in the early universe \citep{Jeon2014}.  Powered by accretion onto black holes or neutron stars, X-ray binaries exhibit strong radiation and a variety forms of outflows \citep{Done2007}. Thus, study of accretion helps us understand how the radiation and outflow are generated, their interactions with the environment, as well as fundamental properties of the central compact object.  

However, the physics with accretion has not been fully understood.  If the accretion rate is low, the accretion flow is believed to be hot and optically thin \citep{Yuan2014}. With a moderate accretion rate, the X-ray spectrum of X-ray binaries can be reasonably described with the standard accretion disk model \citep{Shakura1973}, in which the viscous heat is balanced by local radiation, predicting an optically thick geometrically thin multicolor disk.   In the high accretion regime, i.e., when it is close to or exceeds the rate needed to power the Eddington luminosity, both advection and outflow are expected to take place, and the disk could be highly turbulent.  In this case, no valid analytic models exist to take into account all these issues.  The slim disk model \citep{Abramowicz1988}, which assumes advection instead of radiation to be the dominant cooling mechanism, is found to be stable at high accretion rates and has been used for fitting the energy spectra of luminous X-ray binaries \citep{Watarai2001}.  However, the model is not complete without considering the radiation driven outflow \citep{Poutanen2007}, which has been ubiquitously observed in (ultra)luminous X-ray binaries \citep{Neilsen2009,Middleton2014, Middleton2015, Pinto2016, Kosec2021}.  

The magnetic field should play an essential role in transporting the angular momentum \citep{Balbus1991} and possibly supporting the disk in addition to the thermal and radiation pressure, though usually only the latter two are considered in analytical models. The standard disk model is found to be thermally unstable if the radiation pressure dominates \citep{Shakura1976}, while the slim disk is stable when advection becomes the major cooling term.  \citet{Begelman2007} suggest that the disk could be supported by the magnetic pressure, which saturates when the MRI is sufficiently developed. 

During the outburst of X-ray binaries,  one also needs to assume a hot corona to account for the observed hard, Comptonized X-rays in addition to the soft thermal photons originated in the optically thick multicolor disk. The formation of the corona has been discussed analytically, e.g., a magnetically driven corona \citep{Galeev1979} or a radiation evaporated corona \citep{Meyer1994,Esin1997}. Some observations suggest that the corona is related to the jet base \citep{Markoff2005}. Massive numerical simulations have shown the presence of hot gaseous coronae \citep{MoralesTeixeira2018, Jiang2019, Kinch2020}. Recently, X-ray polarization observations with PolarLight \citep{Long2022} and the Imaging X-ray Polarimetry Explorer \citep{Krawczynski2022} have placed constraints on the corona geometry. Also, in order to incorporate the jet formation, one has to rely on numerical simulations \citep{Davis2020}.

It is challenging to resolve the thin disk with numerical simulations in the subcritical regime. \citet{Hawley2001} and \citet{Hawley2001a} performed global magnetohydrodynamic (MHD) simulations of the thin disk without considering any radiation effect. \citet{Hogg2016,Hogg2018} added an artificial cooling function to approximate the radiation transfer and keep the accretion disk thin.  In the meanwhile,  general relativity magnetohydrodynamic (GR-MHD) simulations \citep{DeVilliers2003, Koide2003, McKinney2004, Shafee2008,Noble2009,Schnittman2013} have been carried out to investigate the accretion flow in the Kerr metric.  The above simulations mainly focused on the estimation of the accretion efficiency and stress-to-pressure ratio of the thin disks. \citet{Ohsuga2006} conducted a 2D radiation hydrodynamic (RHD) simulation of an assumed $\alpha$ disk with radiation transfer taken into account. Their simulation has a large mass input rate of 100 $L_{\rm{Edd}}/c^2$, such that the disk does not always stay in the sub-Eddington thin disk state but displays super-Eddington bursts. Recently, \citet{MoralesTeixeira2018} presented a global simulation for a subcritical thin disk with 3D GR-RMHD codes in the magnetically arrested disk (MAD) state with treatment of radiation transport. \citet{Fragile2018} presented a 2D GR-RHD simulation of viscous Shakura-Sunyaev thin accretion disk around a stellar mass black hole employing M1 scheme for the radiation.

In recent years, more attentions have been paid to the simulation of systems with high accretion rates.  The simulations can be classified into three categories. The 2D simulations can expand to large radii to study the large scale structures of the accretion flow owing to the fewer computational resources that are needed. However, because of the anti-dynamo theorem, the 2D simulations cannot sustain magnetorotational instability (MRI) turbulence self-consistently. As a result, they are either purely RHD with a viscosity assumption \citep{Kawashima2009,Hashizume2015,Kitaki2017,Ogawa2017,Kitaki2018,Kitaki2021} or RMHD with a mean field dynamo approximation \citep{Ohsuga2009,Ohsuga2011,Sadowski2015}.  The 3D GR-RMHD simulations are resource-consuming; they can only resolve the innermost region of the accretion flow, but offer a chance to study the impact of black hole spin on the accretion; they use M1 scheme \citep{Fragile2014,McKinney2014,Takahashi2016,Sadowski2016, Wielgus2022} or variable Eddington tensor \citep{Asahina2022} to handle radiation transport.

Our simulations adopt the psuedo-Newtonian potential but solve the full angular resolved transport equation without assuming the closure relation. Using the same codes, \citet{Jiang2014} performed simulations of a supercritical accretion flow around a stellar mass black hole in cylindrical coordinates. Similar simulations in the spherical coordinates have been done for subcritical \citep{Jiang2019} and supercritical \citep{Jiang2019a} accretion onto a supermassive black hole.  In this paper, we present results from 3D RMHD simulations of accretion flows around a stellar mass black hole in spherical coordinates with different initial conditions, which lead to various accretion rates from sub- to near-critical accretion rates. We try to extract the accretion properties, such as the radiative efficiency, outflow rate, disk structure, and corona temperature, as a function of accretion rate, and analyze the mechanism for angular momentum transfer under different accretion rates.  

The paper is organized as follows. We describe the simulation setup in Section~\ref{sec:setup}. The main features of the accretion flow from simulations are presented in Section~\ref{sec:result}, including the time variation (\ref{subsec:history}), inflow and outflow rates (\ref{subsec:flow_rate}), radiation and advection luminosities (\ref{subsec:luminosity}), 2D disk structure (\ref{subsec:structure}), and 1D radial (\ref{subsec:radial_profile}) and vertical (\ref{subsec:vertical_profile}) disk structures. The results are discussed in Section~\ref{sec:discussion} and summarized in Section~\ref{sec:conclusion}.

\section{Simulation setup}
\label{sec:setup}

We adopt the ideal MHD with radiative transfer in the simulation,  using the same equations as in \citet[][see their Eqs.~1--6]{Jiang2019a}. We carry out the simulations using the code Athena\pp\ \citep{Stone2019} and the method described in \citet{Jiang2021}. We assume the psuedo-Newtonian potential \citep{Paczynsky1980} to mimic the effect of general relativity around a Schwarzschild black hole
 \begin{equation}
\phi = - \frac{G M_{\rm BH}}{r - 2r_{\rm g}} \; ,
\end{equation}
where $G$ is the gravitational constant and $r_{\rm g} \equiv G M_{\rm BH} / c^2$ is the gravitational radius. The Compton scattering effect is treated based on the difference of radiation and gas temperatures. The interactions between gas and radiation are described by a source term in the radiation transport equation, as in \citet[][see their Eqs.~4]{Jiang2019a}.
 
\begin{figure}[tb]
\centering
\includegraphics[width=\columnwidth]{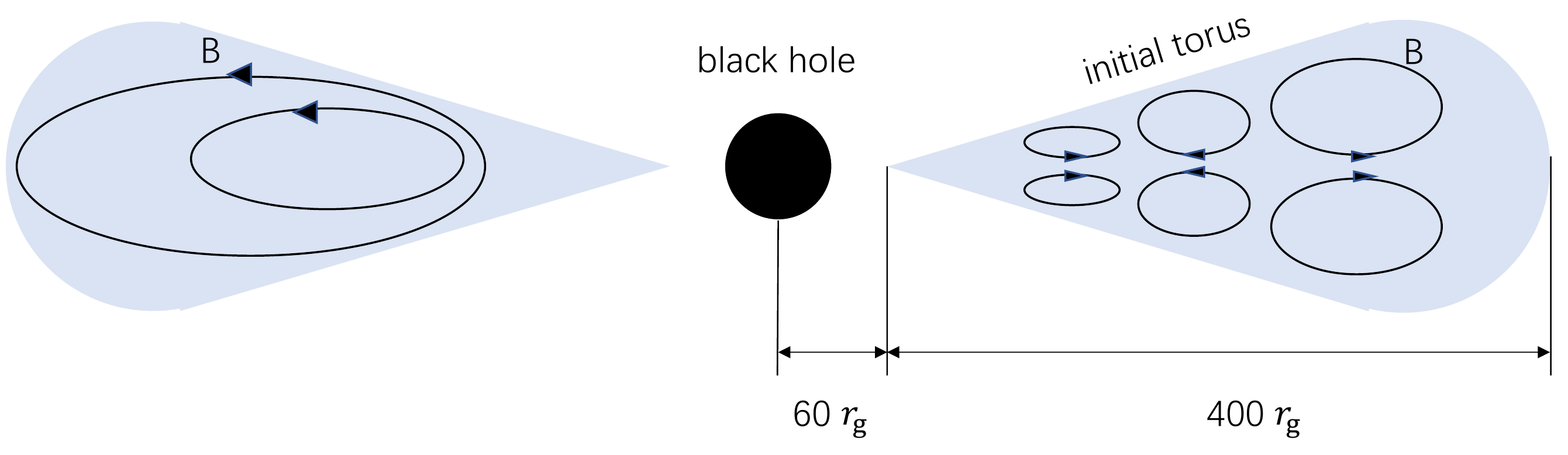}
\caption{Initial torus and magnetic field configuration used in the simulations. The left side shows the initial setup with single-loop magnetic fields (XRB0.9), while the right side shows the case with multiple loops (XRB0.01 and XRB0.8). 
\label{fig:initial_magnetic_field}}
\end{figure}

We carry out three runs of simulations, namely XRB0.01, XRB0.8, and XRB0.9 around a 6.62~$M_\sun$ stellar mass black hole. The fiducial parameters used in the simulation are listed in Table~\ref{tab:fiducial}. We initialize a hydrostatic rotating gas torus with a density maximum at 120~$r_{\rm g}$, with different maximum gas densities and temperatures. The shape of the gas torus is the same as that in \citet{Jiang2019a} and \citet{Jiang2019}. The inner edge of the torus is at 60~$r_{\rm g}$ and we fill the region outside the initial torus with a density floor of $10^{-8}$~$\rho_0$. The initial parameters, including the maximum density $\rho_{\rm i}$, the maximum gas temperature $T_{\rm i}$, and the ratio between the radiation pressure $P_{\rm R}$ and the magnetic pressure $P_{\rm B}$ to the gas pressure $P_{\rm g}$, are summarized in Table~\ref{tab:initial}. The three runs assume different initial magnetic field configurations. Magnetic fields with a single loop are adopted in run XRB0.9, while those with multiple loops are used in runs XRB0.01 and XRB0.8, see Figure~\ref{fig:initial_magnetic_field} for illustration. The different initial parameters and the different magnetic field configurations lead to different mass accretion rates for the disks formed near the central black hole.

The simulation covers the domain of $\left( r, \theta, \phi \right) \in \left( 4 r_{\rm g}, 1600 r_{\rm g} \right) \times \left( 0, \pi \right) \times \left( 0, 2\pi \right)$. The highest resolution reaches $\Delta r/r=\Delta \theta=\Delta \phi=0.012$ near the disk mid-plane. We use 80 discrete angles in each cell to resolve the angular distribution of the radiation field.

\begin{deluxetable*}{cllll}[ht]
\tablecaption{Fiducial Simulation Parameters
\label{tab:fiducial}}
\tablewidth{\textwidth}
\tablecolumns{3}
\tablehead{
\colhead{Parameters} & \colhead{Definition} & \colhead{Value} & \colhead{Physical meaning}
}
\startdata
$M_{\rm BH}$ & $6.62 M_\sun$ & $1.32 \times 10^{34}$ $\rm{g}$ & Black hole mass\\
$r_{\rm g}$ & $G M_{\rm BH} / c^2$ & $9.76 \times 10^{5}$ $\rm{cm}$ & Gravitational radius\\
$\kappa_{\rm es}$ &  & 0.34 $\rm{g}^{-1}$ $\rm{cm}^2$ & Electron scattering opacity\\
$L_{\rm Edd}$ & $4 \pi G M_{\rm BH} c / \kappa_{\rm es}$ & $9.74 \times 10^{38}$ $\rm{erg}$ $\rm{s}^{-1}$ & Eddington luminosity\\
$\dot{M}_{\rm crit}$ & $10 L_{\rm Edd} / c^2$ & $1.08 \times 10^{19}$ $\rm{g}$ $\rm{s}^{-1}$ & Critical accretion rate\\
$\rho_{0}$ & & $10^{-2}$ $\rm{g}$ $\rm{cm}^{-3}$ & Fiducial density\\
$T_{0}$ & & $10^7$ $\rm{K}$ & Fiducial temperature\\
$P_{0}$ & $R_{\rm ideal} \rho_{0} T_{0}$ & $1.39 \times 10^{13}$ $\rm{dyn}$ $\rm{cm}^{-2}$ & Fiducial pressure\\
$v_{0}$ & $\sqrt{R_{\rm ideal} T_{0}}$ & $3.73 \times 10^{7}$ $\rm{cm}$ $\rm{s}^{-1}$ & Fiducial velocity\\
$t_{0}$ & $2 r_g / v_0$ & $5.23 \times 10^{-2}$ $\rm{s}$ & Fiducial time\\
$B_{0}$ & $\sqrt{P_0}$ & $3.73 \times 10^{6}$ $\rm{G}$ & Fiducial magnetic field\\
$E_{0}$ & $a_{\rm R} T_{0}^{4}$ & $7.57 \times 10^{13}$ $\rm{erg}$ $\rm{cm}^{-3}$ & Fiducial radiation energy density\\
$F_{0}$ & $c E_0$ & $2.27 \times 10^{24}$ $\rm{erg}$ $\rm{cm}^{-2}$ $\rm{s}^{-1}$& Fiducial radiation flux\\
\enddata
\tablecomments{We assume an accretion efficiency of 0.1 to relate the critical accretion rate with the Eddington limit. $R_{\rm ideal}$ is the ideal gas constant with a mean molecular weight $\mu = 0.6$. $a_{\rm R} = 7.57 \times 10^{-15}$ $\rm{erg}$ $\rm{cm}^{-3}$ $\rm{K}^{-4}$ is the radiation constant.}
\end{deluxetable*}

\begin{deluxetable}{cccc}[ht]
\tablecaption{Initial Simulation Parameters
\label{tab:initial}}
\tablewidth{\textwidth}
\tablecolumns{3}
\tablehead{
\colhead{Variables/Units} & \colhead{XRB0.01} & \colhead{XRB0.8} & \colhead{XRB0.9}
}
\startdata
$r_{\rm i}/r_{\rm g}$ & 120 & 120 & 120\\
$\rho_{\rm i}/\rho_{0}$ & 0.05 & 6 & 10\\
$T_{\rm i}/T_{0}$ & 1.38 & 4.55 & 5.16\\
$\left< P_{\rm R}/P_{\rm g}\right>$ & $5.22\times10^3$ & 53.6 & 35.6\\
$\left< P_{\rm R}/P_{\rm g}\right>_{\rho}$ & $2.50\times10^2$ & 22.1 & 19.2\\
$\left< P_{\rm B}/P_{\rm g}\right>$ & $2.95\times10^{-2}$ & $2.10\times10^{-4}$ & $1.34\times10^{-3}$\\
$\left< P_{\rm B}/P_{\rm g}\right>_{\rho}$ & $1.18\times10^{-2}$ & $6.22\times10^{-5}$ & $3.97\times10^{-4}$\\
$\Delta r/r$ & 0.012 & 0.012 & 0.012\\
$\Delta \theta$ & 0.012 & 0.012 & 0.012\\
$\Delta \phi$ & 0.012 & 0.012 & 0.012\\
$N_{\rm n}$ & 80 & 80 & 80\\
\enddata
\tablecomments{The center of the initial torus is at $r_{\rm i}$. The initial density and gas temperature at the center of the torus are $\rho_{\rm i}$ and $T_{\rm i}$, respectively. For any quantity $a$, $\left< a \right>$ is the volume-averaged value inside the initial gas torus, and $\left< a \right>_{\rho}$ is the density-weighted averaged value inside the torus. The grid size $\Delta r$, $\Delta \theta$, and $\Delta \phi$ are for the finest grids at the center of the torus. The number of angles for the radiation grid is $N_{\rm n}$ in each cell.}
\end{deluxetable}

\section{Results}
\label{sec:result}

The MRI creates turbulence in the initial mass torus and transports angular momentum outwards. The mass is slowly accreted onto smaller radii and forms an accretion disk self-consistently. Due to the different initial magnetic fields and thus different magnitudes of MRI and angular momentum transfer rates, the three runs lead to distinct mass accretion rates. XRB0.01 has a subcritical ($\sim$$10^{-2}$~$\dot{M}_{\rm crit}$) accretion rate with sub-Eddington emission; XRB0.8 and XRB0.9 are near-critical (0.8--0.9~$\dot{M}_{\rm crit}$)\footnote{Strictly speaking, XRB0.8 and XRB0.9 are also subcritical, but we refer them to be near-critical to distinguish from XRB0.01.} and sub-Eddington. After the initial transition phase, the simulation converges to a quasi-steady state, during which the net accretion rate is relatively constant.  If the standard deviation of net accretion rate during a time span is less than 1/3 of the average, we define this period as the quasi-steady state and perform the analysis in it.

\subsection{Resolution for MRI turbulence}
\label{subsec:resolution}

To determine if the MRI turbulence is adequately built and well-resolved, we calculate the quality factors $Q_{\theta}$ and $Q_{\phi}$ following \citet{Hawley2011} and \citet{Sorathia2012}. The quality factor is defined as the ratio between the fastest growing MRI mode $\lambda = 2 \pi \sqrt{16/15} |v_{\rm A}| / \Omega$ and cell size $r \triangle{\theta}$ or $r \sin{\theta} \triangle{\phi}$, respectively, along $\theta$ or $\phi$, where $v_{\rm A}$ represents the Alfv\'{e}n velocity for $B_{\theta}$ or $B_{\phi}$. The statistical properties of MRI turbulence do not change with grid resolution if $Q_{\theta} \ge 6$, $Q_{\phi} \ge 25$ or both are greater than 10 \citep{Hawley2013}. We regard this as the condition for well-resolved MRI turbulence.

We calculate the azimuthally averaged quality factor in the three runs at radii from 6 to 20~$r_{\rm g}$. For XRB0.01, $Q_{\theta}$ is always larger than 15 near the disk surface and reduces from 8 at 6~$r_{\rm g}$ to 6 at 20~$r_{\rm g}$ near the disk mid-plane. For XRB0.8, $Q_{\theta}$ is found to be greater than 15 everywhere. For XRB0.9, $Q_{\theta}$ reduces from 10 at 6~$r_{\rm g}$ to 6 at 20~$r_{\rm g}$ near disk mid-plane, but is found to be smaller than 6 occasionally in some small regions. $Q_{\phi}$ is over 50 in all the runs. Thus $Q_{\theta} \ge 6$ and $Q_{\phi} \ge 25$ are satisfied in the majority of the central disk, e.g., regions with grid refinement.  

\subsection{Simulation histories}
\label{subsec:history}

\begin{figure}[ht]
\plotone{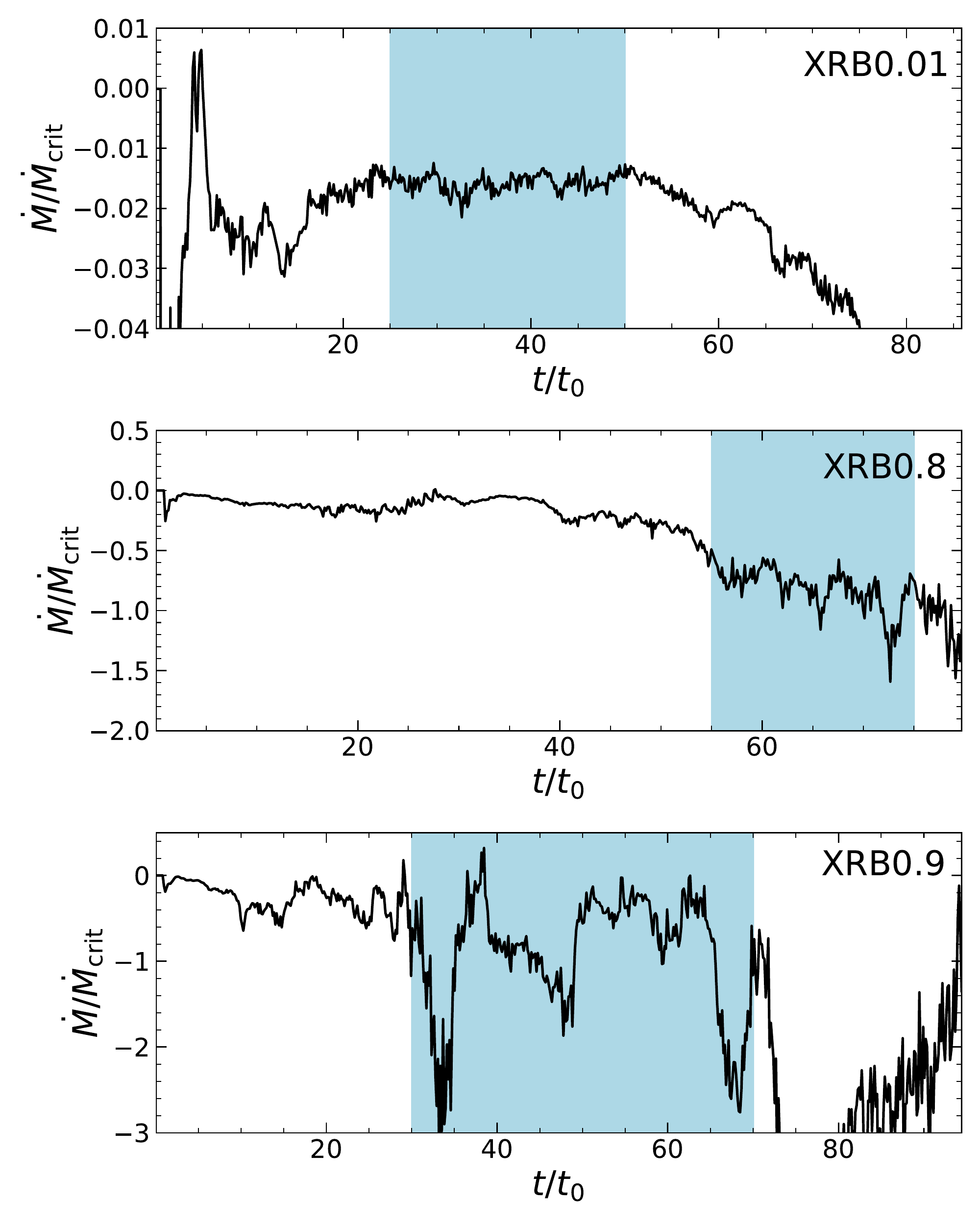}
\caption{Histories of the spherically integrated mass accretion rate in the three runs at 10~$r_{\rm g}$. Negative means gas flows towards the black hole. The quasi-steady states are marked with blue shades. The accretion flow shows no sign of thermal runaway over hundreds of thermal timescales at 10~$r_{\rm g}$.
\label{fig:history}}
\end{figure}

We calculate the net mass accretion rate at radius $r$ as
\begin{equation} 
\dot{M} = \int_{0}^{2 \pi} \int_{0}^{\pi} \rho v_r r^2 \sin{\theta} d\theta d\phi.
\end{equation}
Histories of $\dot{M}$ at 10~$r_{\rm g}$ for the three runs are shown in Figure~\ref{fig:history}. After an initial transition phase of 25--55~$t_0$ in each run, the accretion flows reach a quasi-steady state which lasts for 20--40~$t_0$. We emphasize that the MRI turbulence in the quasi-steady state should be well-resolved. For instance, although the mass accretion rate in XRB0.8 seems steadier in 20--50~$t_0$ than in 55--75 $t_0$, the MRI turbulence quality factors in the former interval are not high enough to justify a quasi-steady state. The Keplerian rotation period of the Paczy\'{n}ski-Wiita potential at radius $r$ is \citep{Jiang2014}
\begin{equation}
t_{\rm K} = 1510 \frac{r_{\rm g}}{c} \left( \frac{r}{40 r_{\rm g}} \right)^{1/2} \left( \frac{r/2 r_{\rm g} - 1}{19} \right) \; .
\end{equation}
Therefore, the duration of each run is equivalent to $\sim$800 orbits at 10~$r_{\rm g}$. Over hundreds of thermal timescales at 10~$r_{\rm g}$,  there is no sign of thermal instability for the three runs. The mass accretion rates in the quasi-steady state have significant fluctuations because of MRI turbulence. The average accretion rates of the three runs are listed in Table~\ref{tab:luminosity}. The standard deviation of the net accretion rate in XRB0.01, XRB0.8, and XRB0.9 is 9.4\%, 21.2\%, and 76.8\%, respectively. In general, the standard deviation in the quasi-steady state is larger when the accretion rate is higher, while XRB0.9 has significantly larger standard deviation due to rapid variations of the accretion rate at times around 33~$t_0$, 48~$t_0$ and 68~$t_0$, which are caused by the variation of the magnetic field strength (see the discussion later in this section). If we omit these time intervals and only consider those with a relatively steady accretion rate, the standard deviation becomes 12.8\% and 53.3\%, respectively, during 40--45~$t_0$ and 51--63~$t_0$.  The fluctuation of accretion rate has a link with the topology of the initial magnetic field.  For the runs XRB0.01 and XRB0.8, quadruple magnetic fields with a net $\bar{B}_r$ component are assumed. The net $\bar{B}_r$ near mid-plane will shear into toroidal magnetic fields and quickly builds up a strong magnetic pressure, which will escape from the mid-plane due to magnetic buoyancy. The MRI turbulence shows less variability in these cases \citep{Pessah2005,Das2018} than in the case where the initial magnetic field has a single loop, i.e., in XRB0.9.

\tabletypesize{\scriptsize}
\begin{deluxetable*}{lcccccccccccccc}
\tablecaption{Mass rates, powers, temperatures, and other key properties of the accretion in the three runs.}
\label{tab:luminosity}
\decimalcolnumbers
\tablewidth{\textwidth}
\tablecolumns{14}
\tablehead{
\colhead{} & \colhead{$\frac{\dot{M}}{\dot{M}_{\rm crit}}$} & \colhead{$\frac{L_{\rm R}}{L_{\rm Edd}}$} & \colhead{$\frac{L_{\rm k}}{L_{\rm Edd}}$} & \colhead{$\frac{L_{\rm R, BH}}{L_{\rm Edd}}$} & \colhead{$\frac{L_{\rm k, BH}}{L_{\rm Edd}}$} & \colhead{$\frac{\dot{M}_{\rm w}}{\dot{M}_{\rm crit}}$} & \colhead{$\eta_{\rm R}$} & \colhead{$\eta_{\rm R, BH}$} & \colhead{$\xi_{\rm w}$} & \colhead{$\theta_{\rm d}$} & \colhead{$T_{\rm c}$} & \colhead{$T_{\rm ph}$} & \colhead{$\frac{T_{\rm axis}}{\rm K}$} & \colhead{$\frac{v_{\rm w}}{c}$} 
}
\startdata
XRB0.01 & $-$0.0158 & 0.0096 & 0 & $-$0.00068 & $-$0.034 & 0 & 6.1\% & 0.4\% & 0 & $20^{\circ}$ & $r^{-0.57}$ & $r^{-0.59}$ & $3\times10^9$ & \\
XRB0.8 & $-$0.82 & 0.22 & 0.0023 & $-$0.085 & $-$1.6 & 0.021 & 2.6\% & 1.0\% & 2.5\% & $56^{\circ}$ & $r^{-0.38}$ & $r^{-0.44}$ & $2\times10^9$ &  0.122\\
XRB0.9 & $-$0.9 & 0.34 & 0.0059 & $-$0.091 & $-$2.3 & 0.025 & 3.7\% & 1.0\% & 2.7\% & $73^{\circ}$ & $r^{-0.48}$ & $r^{-0.29}$ & $8\times10^8$ & 0.096 \\
\enddata
\tablecomments{
Column~1: Run name.
Column~2: Normalized net mass accretion rate.
Column~3: Normalized radiation luminosity. 
Column~4: Normalized kinematic luminosity.
Column~5: Normalized radiation luminosity swallowed by the central black hole.
Column~6: Normalized kinematic luminosity swallowed by the central black hole.
Column~7: Normalized wind mass loss rate.
Column~8: Efficiency of radiation.
Column~9: Efficiency of swallowed radiation.
Column~10: Ratio of the true outflow mass rate to the net accretion rate.
Column~11: Half open angle of the effective absorption photosphere, or the central low-density funnel, measured from the disk mid-plane.
Column~12: Mid-disk temperature as a function of radius.
Column~13: Radial temperature profile on the effective absorption photosphere.
Column~14: Corona temperature.
Column~15: Maximum wind/outflow velocity. 
}
\end{deluxetable*}
\tabletypesize{\normalsize}

\begin{figure*}[ht]
\centering
\includegraphics[width=1.0\textwidth]{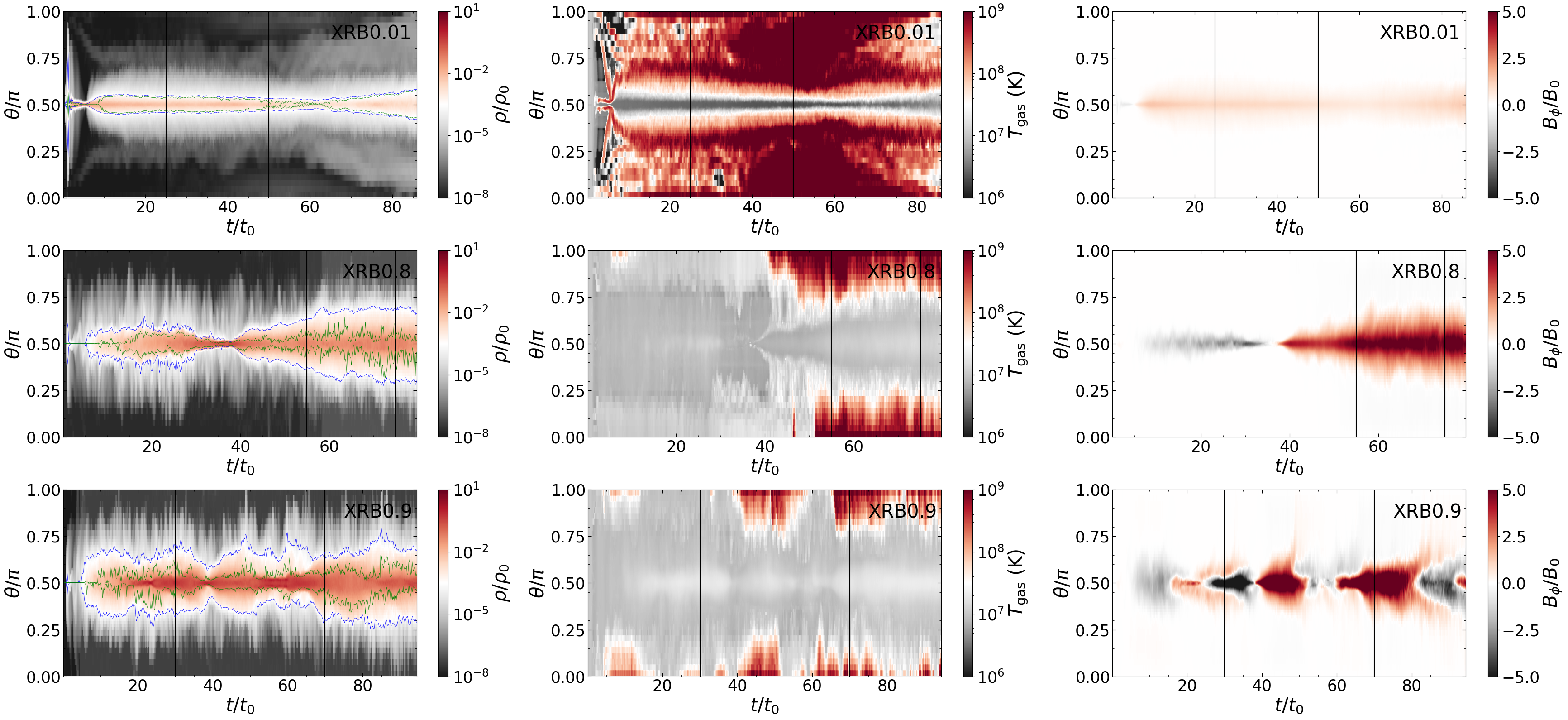}
\caption{Space-time diagrams of azimuthally averaged density (left column), gas temperature (middle column) and azimuthal magnetic field component (right column) at 20~$r_{\rm g}$ for the three runs. The two black vertical lines in each panel indicate the start and end of the quasi-steady state, respectively (see Figure \ref{fig:history}). The green and blue lines in the density diagrams indicate the position of effective absorption and electron scattering photospheres, respectively, measured from the rotational axis.
\label{fig:space_time}}
\end{figure*}

Histories of the azimuthally averaged profiles (space-time diagrams) as a function of $\theta$ for the three runs are displayed in Figure~\ref{fig:space_time}, for the density $\rho$, gas temperature $T$, and azimuthal magnetic field component $B_{\phi}$ at 20~$r_{\rm g}$. When the simulation enters the quasi-steady state at 20~$r_{\rm g}$ (marked between two vertical black lines), there is a clear positive correlation between the disk scale height and mass accretion rate. The photosphere for effective absorption (green curves) and electron scattering (blue curves), which are integrated from the rotational axis, are shown on top of the density profile; their heights are also scaled with accretion rate. High temperature coronae ($10^8-10^9$~K) can be seen above the electron scattering photosphere, consistent with the previous global simulation \citep{Jiang2019,Jiang2014} and local shearing box simulation \citep{Jiang2014a}. For the run XRB0.9 that has single loop magnetic fields initially, $B_{\phi}$ repeatedly flips its direction near the disk mid-plane every 10~$t_0$ after the disk enters the quasi-steady state, which is roughly $\sim$30 Keplerian rotation periods at 20~$r_{\rm g}$. The magnetic buoyancy drives matter with a strong magnetic field up from the disk mid-plane, forming the so-called butterfly diagram. The butterfly diagram has been observed in previous global simulations \citep{Jiang2014,Jiang2019a,Jiang2019} and local shearing box simulations \citep{Stone1996,Miller2000,Davis2010,Shi2010,Simon2012,Jiang2013,Jiang2014a, Salvesen2016, Salvesen2016a}, believed to be related with a dynamo process of the MRI \citep{Brandenburg1995,Blackman2012}. The butterfly diagram is not seen in runs XRB0.01 and XRB0.8 that have quadruple magnetic fields initially, because their magnetic fields have a net radial component near the disk mid-plane, and this component always transfers to $B_{\phi}$ in the same direction by MRI.

In XRB0.9, when the magnetic field flips its sign and reaches a minimum strength, the disk shrinks to a small scale height, suggesting that the disk is magnetic pressure supported in this near-critical case. Details about the disk pressure will be discussed in Sections \ref{subsec:radial_profile} and \ref{subsec:vertical_profile}. Meanwhile, when the magnetic field strength reaches a maximum value, the angular momentum transport of MRI turbulence is enhanced and the accretion rate shows a sudden increase (see Figure \ref{fig:history} and Figure \ref{fig:space_time}).

\subsection{Inflow and outflow rate}
\label{subsec:flow_rate}

\begin{figure*}[ht]
\centering
\includegraphics[width=1.0\textwidth]{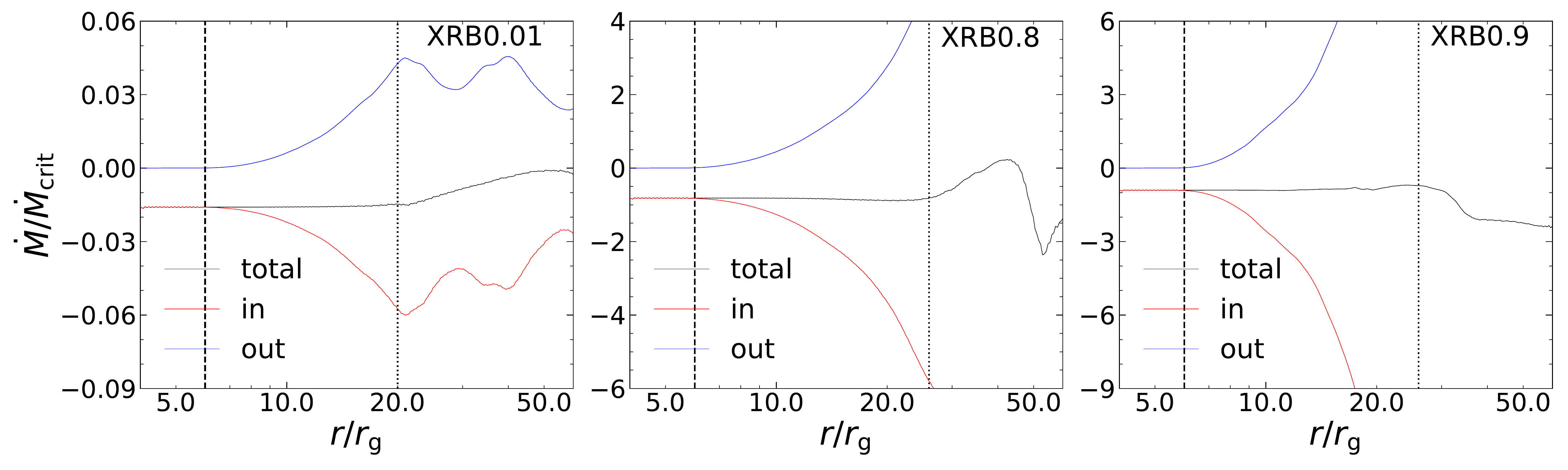}
\caption{Time-averaged radial profiles of the mass accretion rate normalized to the critical rate. In each panel, the solid black lines are the net mass accretion rates, the solid blue and red lines are the mass outflow and inflow rates, respectively. The black dashed lines indicate the location of ISCO (6~$r_{\rm g}$). The black dotted lines indicate the outer radius of the quasi-steady disk. 
\label{fig:radial_profile}}
\end{figure*}

We calculate the time-averaged net mass accretion rate as a function of radius using the following equation,
\begin{equation} 
\label{equ:M_dot}
\langle \dot{M} \rangle = \frac{1}{\Delta{t}} \int_{0}^{2 \pi} \int_{0}^{\pi} \int_{t_1}^{t_2} \rho v_r r^2 dt \sin{\theta} d\theta d\phi \; ,
\end{equation}
where $\Delta{t} = t_2 - t_1$ is the time duration of the quasi-steady state. The radial profiles for the three runs are shown in Figure~\ref{fig:radial_profile}. The net mass accretion rate keeps roughly constant up to 20~$r_{\rm g}$ for the run XRB0.01 and up to 26~$r_{\rm g}$ for XRB0.8 and XRB0.9; these are the radial ranges where the quasi-steady state is obtained. According to Eq.~(\ref{equ:M_dot}), we are averaging over cells with both inward and outward-moving gases. Therefore, the net mass accretion rate can be divided into the two components. We calculate the mass outflow rate by integrating cells with $v_r > 0$ and the mass inflow rate over those with $v_r < 0$. The mass outflow and inflow rates are also shown in Figure~\ref{fig:radial_profile} with blue and red curves. We note that the outflow defined in this way contains true outflows that will escape to infinity, failed outflows \citep{Kitaki2021} that will eventually fall back onto the accretion disk at larger radii, and turbulence inside the accretion disk. The turbulent fluctuations are the dominant component, more than an order of magnitude higher than the other two components. Both the mass outflow and inflow rates are much larger than the net mass accretion rate $\dot{M}$. The outgoing gas emerges outside the innermost stable circular orbit (ISCO) and the mass loss rate increases quickly with radius. We find the mass inflow and outflow rate are significantly smaller than in the AGN case as expected by \citet{Jiang2019a}. Note that our simulation does not take into account the effects of general relativity thus the central black hole is non-rotating. Therefore, the outflows seen near the ISCO are driven by the radiative and magnetic force but with no contribution from the Blandford-Znajec (BZ) mechanism \citep{Blandford1977}.

\subsection{Luminosity and advection}
\label{subsec:luminosity}

To estimate the total radiative flux and the kinetic energy carried away by the true outflow, which is the outflow that can escape to infinity, we make the integration through a cylindrical surface. The radius of the surface is set as the outer radius of the quasi-steady disk, such that the accretion flow within it has reached the quasi-steady state.  The total radiative luminosity $L_{\rm R}$, kinetic luminosity $L_{\rm k}$, and true outflow mass flux $\dot{M}_{\rm w}$ are calculated as
\begin{equation}
\begin{aligned}
&L_{\rm R} = \int_{0}^{r_0} 2 \pi F_{{\rm R}, z} r dr + \int_{-z_0}^{z_0} 2 \pi F_{{\rm R}, r} r_0 dz \; ,\\
&L_{\rm k} = \int_{0}^{r_0} 2 \pi v_z \left( \frac{1}{2} \rho v^2 \right) r dr + \int_{-z_0}^{z_0} 2 \pi v_r \left( \frac{1}{2} \rho v^2 \right) r_0 dz \; ,\\
&\dot{M}_{\rm w} = \int_{0}^{r_0} 2 \pi \rho v_z r dr + \int_{-z_0}^{z_0} 2 \pi \rho v_r r_0 dz \; ,
\end{aligned}
\end{equation}
where $r_0$ and $z_0$ are the radius and half height of the cylindrical surface, and $F_{\rm R}$ is the radiation flux. We integrate through both the upper and lower sides of the disk. Only positive $F_{{\rm R}, r}$ and $v_r$ are considered in the integration in order to exclude the inflow component of the turbulent disk. For $L_{\rm k}$ and $\dot{M}_{\rm w}$, we only include cells where the sum of kinetic and gravitational energy is positive, which implies that the gas will escape to infinity and be a true outflow. The time-average is done during the quasi-steady state.

To find the suitable half height for the integration, we raise $z_0$ to see how it affects the radiative and kinetic luminosity. The radiative luminosity $L_{\rm R}$ saturates at 400--1000~$r_{\rm g}$, while the kinetic luminosity $L_{\rm k}$ saturates at 1500~$r_{\rm g}$ for the three runs. Thus, these $z_0$ are used for integration. The radiation and outflows at these heights are mainly coming from the innermost quasi-steady area, according to the large scale stream lines of velocity and radiation flux. The luminosities calculated in this section only represent the lower limits of their true values, because the integrated radiative and mechanical luminosities for these runs increase exponentially with radius beyond the quasi-steady region. In the wind, radiative and internal energy may convert to mechanical energy and drive more gases to escape to infinity. Thus, we also try to include cells where the total energy $E_{\rm t} = \frac{1}{2} \rho v^2 + \frac{\gamma P}{\gamma - 1} + \rho \phi + \frac{E_{\rm R}}{3}$ is higher than zero. $L_{\rm k}$ and $\dot{M}_{\rm w}$ will increase by a factor of about 2 if we consider the possible conversion of energy. On the contrary, an inverse conversion of energy may lower $L_{\rm k}$ and $\dot{M}_{\rm w}$. We define the lower limit of the radiative efficiencies as $\eta_{\rm R} = -L_{\rm R} / (\dot{M} c^2)$. The ratio of true outflow rate to net accretion rate that goes through ISCO is calculated as $\xi_{\rm w} = -\dot{M}_{\rm w} / \dot{M}$. The luminosities and efficiencies of the three runs are listed in Table~\ref{tab:luminosity}.

Advection of radiative and kinetic energy is an important cooling mechanism for accretion flows with high accretion rates. To evaluate the level of advection, we calculate the radiative energy $L_{\rm R, BH}$ and kinetic energy $L_{\rm k, BH}$ that are swallowed by the central black hole. The integration is computed through a spherical surface at ISCO as
\begin{equation}
\begin{aligned}
&L_{\rm R, BH} = \int_{0}^{2 \pi} \int_{0}^{\pi} F_{{\rm R}, r} \left( 6 r_{\rm g} \right) ^{2} \sin{\theta} d\theta d\phi \; , \; {\rm and}\\
&L_{\rm k, BH} = \int_{0}^{2 \pi} \int_{0}^{\pi} v_r \left( \frac{1}{2} \rho v^2 \right) \left( 6 r_{\rm g} \right) ^{2} \sin{\theta} d\theta d\phi \; ,\\
\end{aligned}
\end{equation}
over cells with negative $F_{{\rm R}, r}$ and $v_r$ to only include the inflow part. The swallowed radiative fraction of the accretion flow is defined as $\eta_{\rm R, BH} = L_{\rm R, BH} / (\dot{M} c^2)$. The swallowed power and its fraction in the three runs are listed in Table~\ref{tab:luminosity}.

The radiative efficiencies for outward radiation calculated in the three runs are around a few percent, comparable to the values reported in \citet{Jiang2014,Jiang2019a}. When the mass accretion rate approaches the critical value, the advection of radiative energy becomes more important; $\eta_{\rm R, BH}$ rises from 0.4\% in the sub-critical case to 1.0\% in the near-critical case. Correspondingly, the radiative efficiency drops from 6.1\% to $\sim$3\% since more radiative energy is swallowed by the black hole. The drop of radiative efficiency with increasing mass accretion rate is a result of increasing advection and outflows at the same time ($\xi_{\rm w}$ increases slightly from XRB0.8 to XRB0.9). Since the radiation pressure increases with increasing accretion rate, more gases are lost via true outflows. In XRB0.01, where the mass accretion rate is the lowest among the three runs, no true outflows can be detected.  On the other hand, the efficiency of swallowed kinetic power is always around 20\%, with a weak correlation with accretion rate.

\subsection{Spatial structure of the disk}
\label{subsec:structure}

\begin{figure*}
\centering
\includegraphics[width=0.8\textwidth]{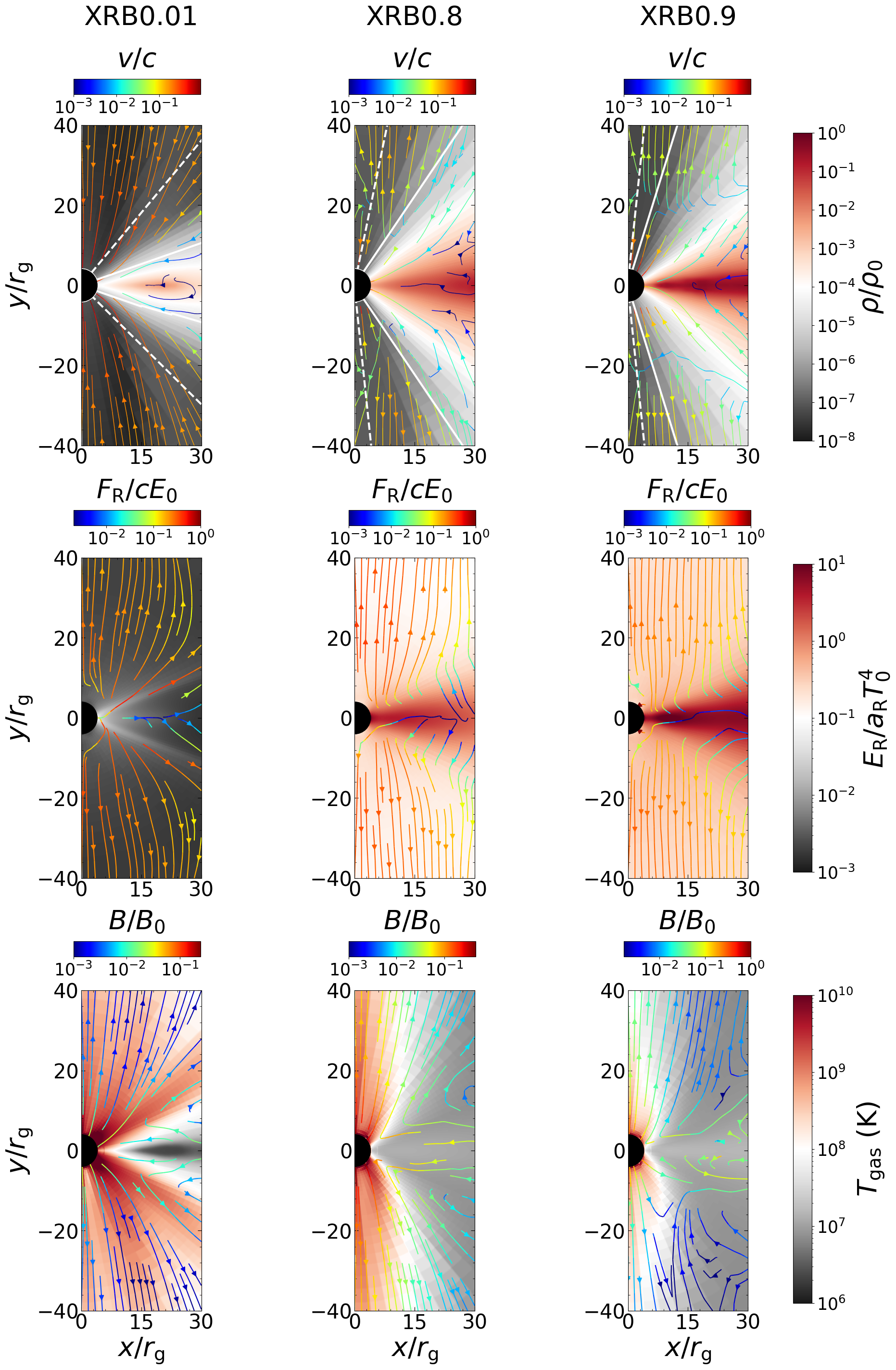}
\caption{Time and azimuthally averaged spatial structures of the accretion flow in the three runs. {\bf Top row}: density (color maps) and mass-weighted flow velocity (streamlines). The white solid and dashed lines represent the photosphere for effective absorption and electron scattering, respectively. {\bf Second row}: radiation energy (color maps) and radiation flux (streamlines). {\bf Third row}: gas temperature (color maps) and magnetic fields (streamlines).
\label{fig:disk_structure}}
\end{figure*}

We calculate the time and azimuthally averaged distributions of the density $\rho$, radiation energy $E_{\rm R}$ and gas temperature $T_{\rm gas}$ in the inner region of the disk, and overlay them with the streamlines of density weighted flow velocity, radiation flux, and magnetic fields, respectively in Figure~\ref{fig:disk_structure}. We also calculate the electron scattering and effective absorption optical depth radially from the outer edge (about 1600~$r_{\rm g}$) of the simulation box, and identify the photosphere locations where the optical depth reaches unity. Note that here the definition is different from that in Section~\ref{subsec:history}, where the integration starts from the rotational axis. 

The accretion flow near the disk mid-plane is dominated by inflows in all runs, while strong outflows are formed inside the low-density funnel (regions encircled by the effective absorption photosphere) with a velocity of $\sim$$0.1c$ except for the low accretion rate case XRB0.01 that has no true outflow. We define the point where the radial velocity $v_r$ changes its sign at the axis as the stagnation point. The stagnation points where outflows are launched are located at about 20~$r_{\rm g}$, which is similar in the super-Eddington AGN case \citep{Jiang2019a}. The disk becomes thicker as the accretion rate increases, which results in a narrower funnel. The effective absorption photosphere also thickens with increasing accretion rate; its half opening angle measured from the disk mid-plane increases from 20\arcdeg\ to 70\arcdeg\ when the accretion rate rises from sub- to near-critical.  As already mentioned in Section \ref{subsec:history}, high temperature coronae can be seen above the effective absorption photosphere inside the low-density funnel region. The temperature of coronae are roughly anti-correlated with the accretion rate. The corona size shrinks when the accretion disk thickens as a result of narrower funnels.

\begin{figure}
\centering
\includegraphics[width=0.45\textwidth]{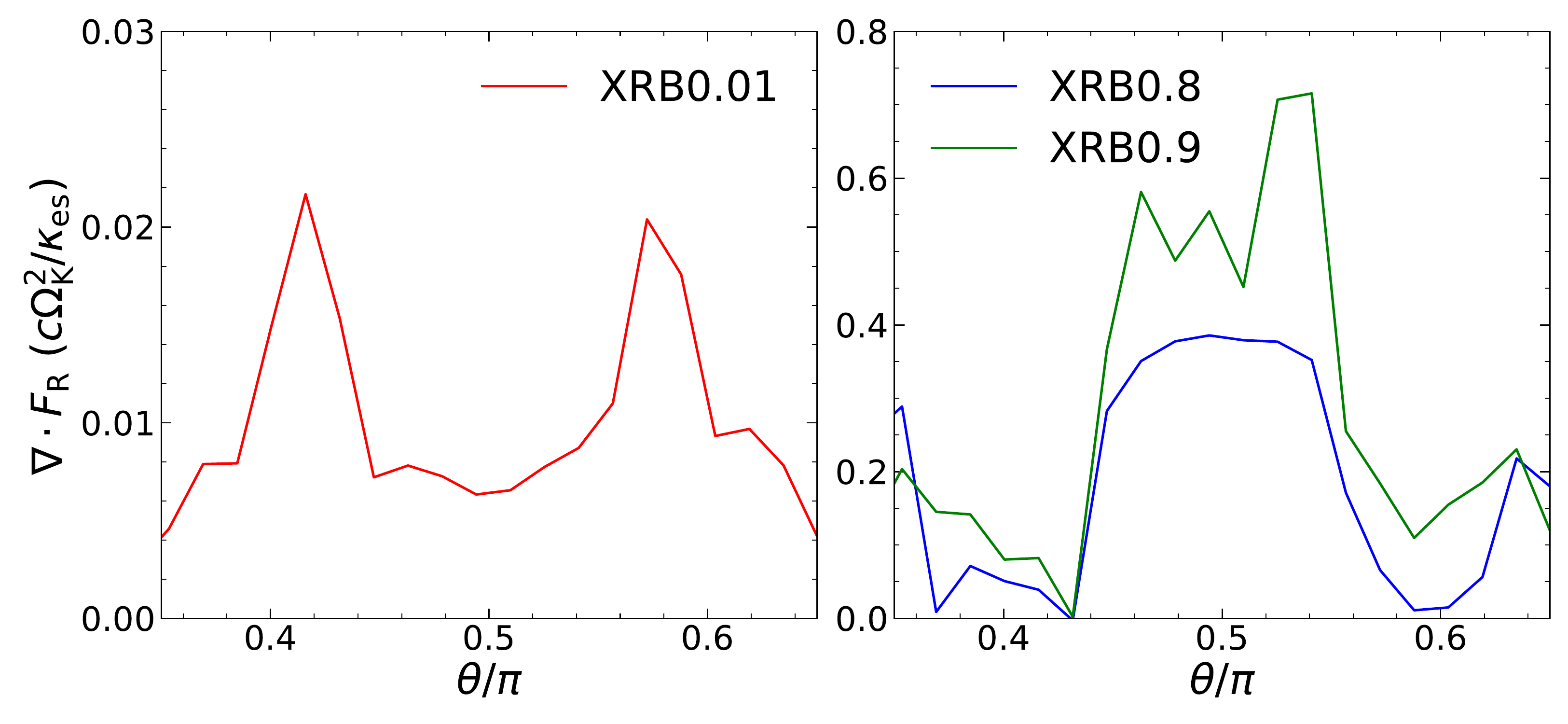}
\caption{Time-averaged vertical profiles of the radiation flux divergence at 10~$r_{\rm g}$ for the three runs.
\label{fig:dissipation}}
\end{figure}

The radiation energy density peaks near the disk mid-plane. However, in the case XRB0.01 where the accretion rate is low, it has a relatively lower density in the mid-plane but peaks at the disk surfaces. Similar distributions have been seen in previous simulations \citep{Jiang2019}. In the runs XRB0.8 and XRB0.9, where the radiation luminosity is $\sim$0.1~$L_{\rm Edd}$, the lab frame radiation flux inside the disk is dominated by the advection term $v_rE_{\rm R}$, while above the disk region the flux flows out roughly vertically at its local radius. In the low accretion rate run XRB0.01, the photons flows out nearly radially through the low density funnel. We show the vertical profiles of the radiation flux divergence in Figure~\ref{fig:dissipation}, to examine the location of the energy dissipation. The energy dissipation is enhanced near the disk surface in the run XRB0.01, but near the disk mid-plane in the other two. In other words, the energy dissipation mainly happens inside the disk when the accretion rate is close to the critical value, but near the disk surface when the accretion rate is low. The energy dissipation is not uniform vertically inside the disk in all the three runs, which is different from the assumption of the \citet{Shakura1973} model  that the dissipation rate is independent of the distance from the mid-plane.

The magnetic field structure is determined by the initial magnetic field topology. In the runs with multiple magnetic field loops (XRB0.01 and XRB0.8) initially, net radial magnetic field near the disk mid-plane are induced. In the run with a single magnetic field loop (XRB0.9) initially, net poloidal magnetic fields are produced and thread through the disk. However, we emphasize both magnetic fields are able to produce magnetic pressure supported disks. The details of disk pressure profile will be discussed below in Section \ref{subsec:radial_profile} and Section \ref{subsec:vertical_profile}.

\subsection{Radial profiles of the disk}
\label{subsec:radial_profile}

We calculate the time-averaged radial profile of any quantity $a$ as
\begin{equation} 
\left< \left< a \right> \right> = \frac{\int_{0}^{2 \pi} \int_{\theta_1}^{\theta_2} \int_{t_1}^{t_2} a dt \sin{\theta} d\theta d\phi}{\Delta{t} \int_{0}^{2 \pi} \int_{\theta_1}^{\theta_2} \sin{\theta} d\theta d\phi},
\end{equation}
where $\Delta{t} = t_2 - t_1$ is the time duration of the quasi-steady state, and $\theta_{1,2}$ correspond to the range of bound gas, which has a total energy $E_{\rm t} = \frac{1}{2} \rho v^2 + \frac{\gamma P}{\gamma - 1} + \rho \phi + \frac{E_{\rm R}}{3}$ lower than 0. The time- and mass-weighted radial profile of any quantity $a$ is defined as
\begin{equation} 
\left< \left< a \right> \right>_{\rho} = \frac{\int_{0}^{2 \pi} \int_{\theta_1}^{\theta_2} \int_{t_1}^{t_2} a \rho dt \sin{\theta} d\theta d\phi}{\Delta{t} \int_{0}^{2 \pi} \int_{\theta_1}^{\theta_2} \rho \sin{\theta} d\theta d\phi}.
\end{equation}

\begin{figure*}[ht]
\centering
\includegraphics[width=1.0\textwidth]{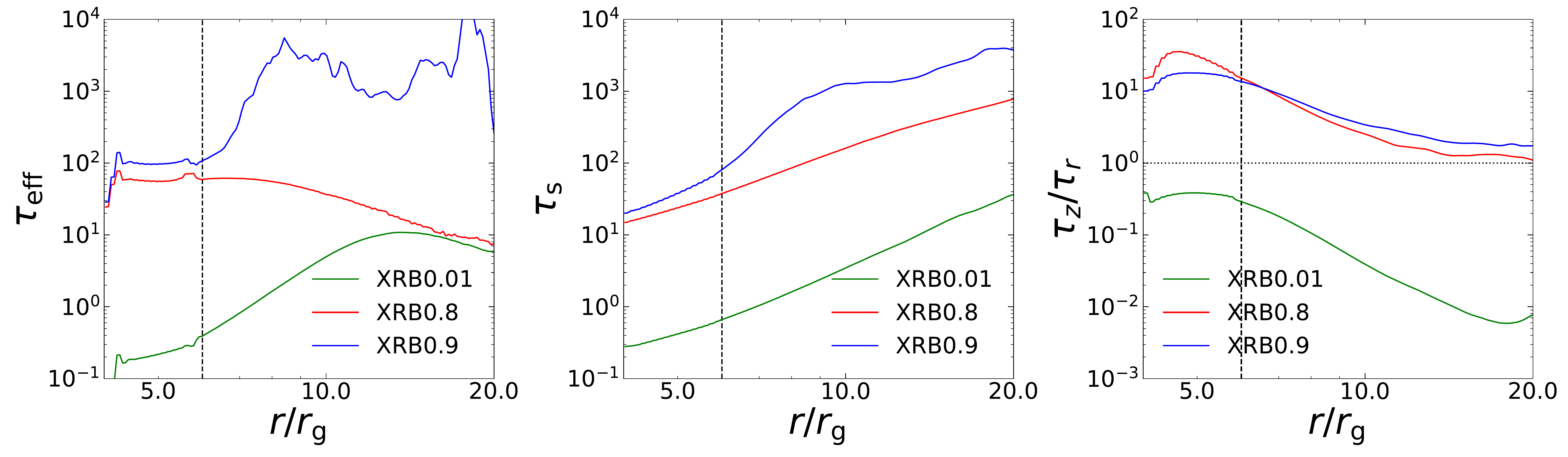}
\caption{Time-averaged radial profile of vertical effective optical depth (left), electron scattering optical depth (middle), and the ratio between the vertical escape timescale and radial advection timescale (right) for the three runs. The vertical dashed line marks the position of the ISCO at 6~$r_{\rm g}$. The horizontal dotted line marks the place where the vertical escape timescale is comparable to the radial advection timescale (i.e. the location where photon trapping is important).
\label{fig:photon_trapping}}
\end{figure*}

First, we show the radial profiles of vertical effective optical depth and electron scattering optical depth in the left two panels of Figure \ref{fig:photon_trapping}. The vertical optical depth at a specific radius $r$ is defined as $\tau=\int_{0}^{\pi} \kappa \rho r d\theta$, where $\kappa$ is the effective or electron scattering opacity. The optical depth increases with increasing disk radius (except for the effective optical depth of XRB 0.8) because of larger surface density, and increases with increasing net mass accretion rate. The optical depth is larger than $10^2$ measured vertically when the accretion rate is close to the critical value in the runs XRB0.8 and XRB0.9 and is $\sim$10 in the low accretion rate run XRB0.01. Except for the innermost region near the ISCO in the run XRB0.01, the disks are always optically thick vertically. We also compare the radial advection timescale of the accretion disk, $\tau_r = r / v_r$, with the estimated vertical escape timescale, $\tau_z = H / v_{{\rm tran}, z}$, for the three runs XRB0.01, XRB0.8, and XRB0.9 within $10^{\circ}$ of the disk mid-plane, where $H$ is the scale height of the effective absorption photosphere, $v_{\rm tran}=F_{\rm R}/E_{\rm R}$ is the effective energy transport speed, and $v_{{\rm tran}, z}$ is the vertical component. The ratio of $\tau_z/\tau_r$ is shown in the right panel of Figure \ref{fig:photon_trapping}. When the vertical escape timescale exceeds the radial advection timescale, photons will be trapped with the inward gas flow. We note here the $v_{\rm tran}$ is actually higher than the photon diffusion speed because of the existence of other transport mechanisms, such as magnetic buoyancy. The photon trapping effect becomes important when the mass accretion rate is close to the critical rate as in runs XRB0.8 and XRB0.9, and this effect is greatly enhanced in disk regions inside 10~$r_{\rm g}$. The photon trapping is not important in the run XRB0.01.

\begin{figure*}[ht]
\centering
\includegraphics[width=\textwidth]{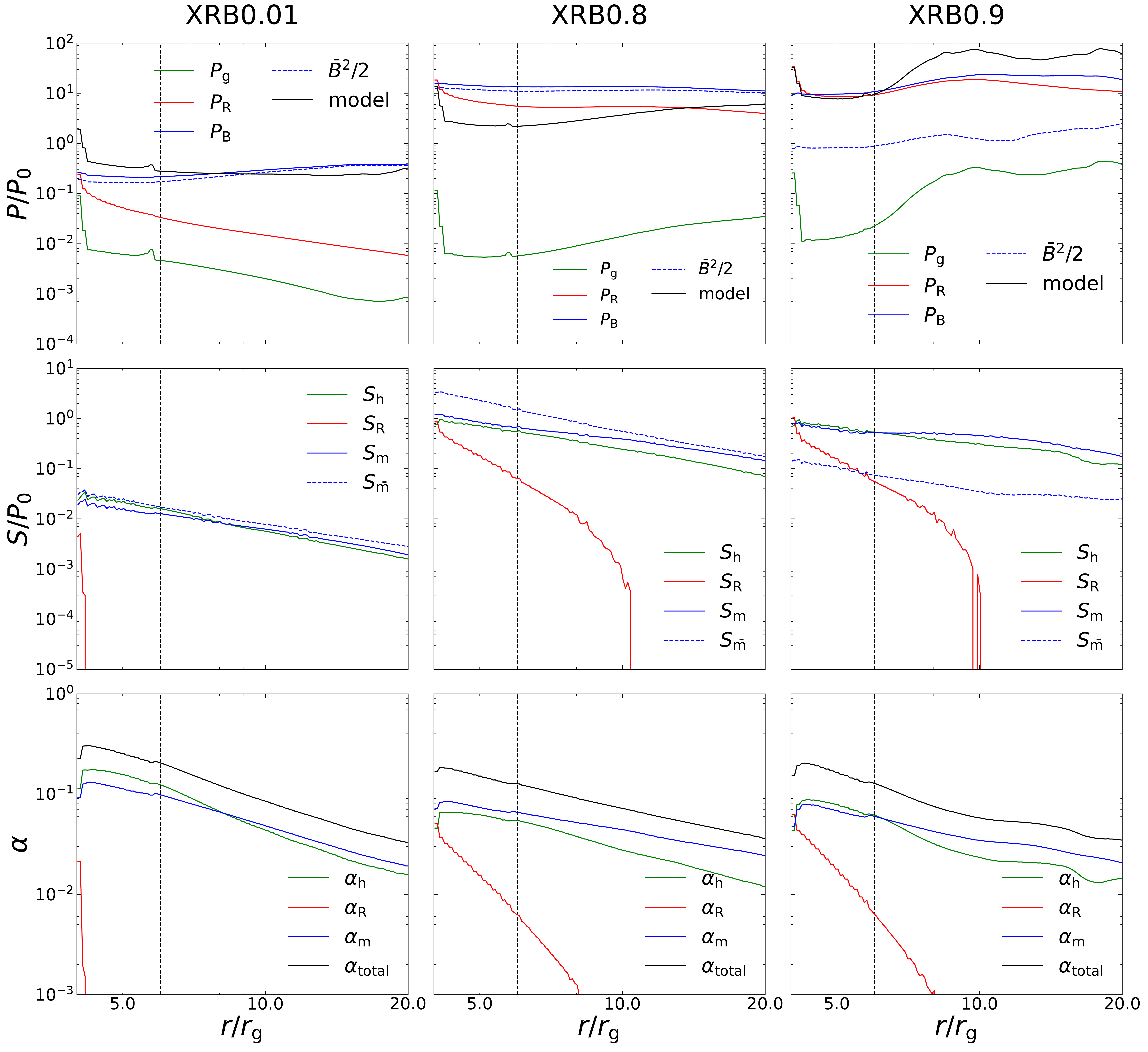}
\caption{Time-averaged radial profile of pressure, stress and effective $\alpha$ for the three runs.  \textbf{Top row}: mass-weighted radial profile of gas (green), radiation (red) and magnetic (blue) pressures. The blue dashed line shows the magnetic pressure due to the mean magnetic field. The black solid line indicates the magnetic pressure quoted from \citet{Begelman2007}. \textbf{Middle row}: radial profile of Reynolds (green), radiation (red) and Maxwell (blue) stresses. The blue dashed line shows the Maxwell stress of the mean magnetic field. \textbf{Bottom row}: radial profile of the effective $\alpha$ parameter due to Reynolds (green), radiation (red), Maxwell (blue) stresses and the total (black). The vertical dashed line marks the position of the ISCO at 6~$r_{\rm g}$.
\label{fig:radial_structure}}
\end{figure*}

We show the mass-weighted radial profiles of the gas pressure $P_{\rm g}$, isotropic radiation pressure $P_{\rm R} = E_{\rm R} / 3$, and magnetic pressure $P_{\rm B}$ in Figure~\ref{fig:radial_structure}. The magnetic pressure is comparable to or dominant over the radiation pressure in the range from ISCO to 20~$r_{\rm g}$. This is similar to the simulation result for a supermassive black hole \citep{Jiang2019}. The gas pressure is 2--3 orders of magnitude lower than the radiation pressure in all runs, consistent with a previous cylindrical simulation \citep{Jiang2014}. We also show the magnetic pressure contributed by the time-averaged magnetic field over the quasi-steady period, i.e., the non-turbulent component (the blue dashed lines in Figure~\ref{fig:radial_structure}).  In the run initially with net poloidal magnetic fields (XRB0.9 with single-loop magnetic fields), the non-turbulent magnetic pressure contributes less than 10\%, suggesting that the magnetic pressure is dominated by the turbulent component. For the other two runs (XRB0.01 and XRB0.8 with multi-loop magnetic fields), the total magnetic pressure is dominated by the non-turbulent component.

Radial profiles of the stress that may account for angular momentum transfer are also shown in Figure~\ref{fig:radial_structure}. We calculate the turbulent component of Maxwell stress $S_{\rm m} = \left< \left< - B_x B_{\phi} \right> \right> + \left< \left< B_x \right> \right> \left< \left< B_{\phi} \right> \right>$, where $B_x = B_r \sin{\theta} + B_{\theta} \cos{\theta}$; the mean magnetic field component of Maxwell stress $S_{\bar{\rm m}} = - \left< \left< B_x \right> \right> \left< \left< B_{\phi} \right> \right>$; and the Reynolds stress $S_{\rm h} = \left< \left< \rho v_x v_{\phi} \right> \right> - \left< \left< \rho v_x \right> \right> \left< \left< v_{\phi} \right> \right>$, where $v_x = v_r \sin{\theta} + v_{\theta} \cos{\theta}$. Here, the angular momentum carried by the mean inflow in the accretion disk is subtracted for the Reynolds stress. We have $S_{\rm R} = \left< \left< P_{\rm R}^{r \phi} \sin{\theta} + P_{\rm R}^{\theta \phi} \cos{\theta} \right> \right>$ for radiation stress. The Maxwell stress is slightly higher than the Reynolds stress for angular momentum transfer and its magnitude during the quasi-steady state is scaled with the vertical component of the magnetic flux, see Figure~\ref{fig:disk_structure}. Similar phenomena have been found in various simulations \citep{Hawley1995, Bai2013, Fromang2013, Simon2013, Bethune2017, Zhu2018, Jiang2019a}. In the run XRB0.9 with net poloidal magnetic field, the turbulent component of Maxwell stress is larger than the mean component by an order of magnitude. For the other two runs XRB0.01 and XRB0.8, there are large mean azimuthal magnetic fields because of shearing of initial radial fields near the disk mid-plane. So the mean field component of Maxwell stress is larger than or comparable to the turbulent component. The radiation stress plays an unimportant role in all three runs, in particular when the accretion rate is quite low, which is in contrast with the AGN simulation \citep{Jiang2019}.

The effective $\alpha$ parameter as a function of radius is shown on the bottom row in Figure~\ref{fig:radial_structure}. The $\alpha$ parameter is found in the range of $\sim$0.03--0.2, and has a similar value in the three runs. This is similar to the previous simulation in cylindrical coordinates \citep{Jiang2014}. The radiation stress has a negligible contribution to the effective $\alpha$ because the disk is optically thick and the mean free path of photons is small. As a result, the anisotropic component of the radiation field needed to produce angular momentum transfer is greatly suppressed.

\subsection{Vertical profiles of the disk}
\label{subsec:vertical_profile}

\begin{figure*}[ht]
\centering
\includegraphics[width=1.0\textwidth]{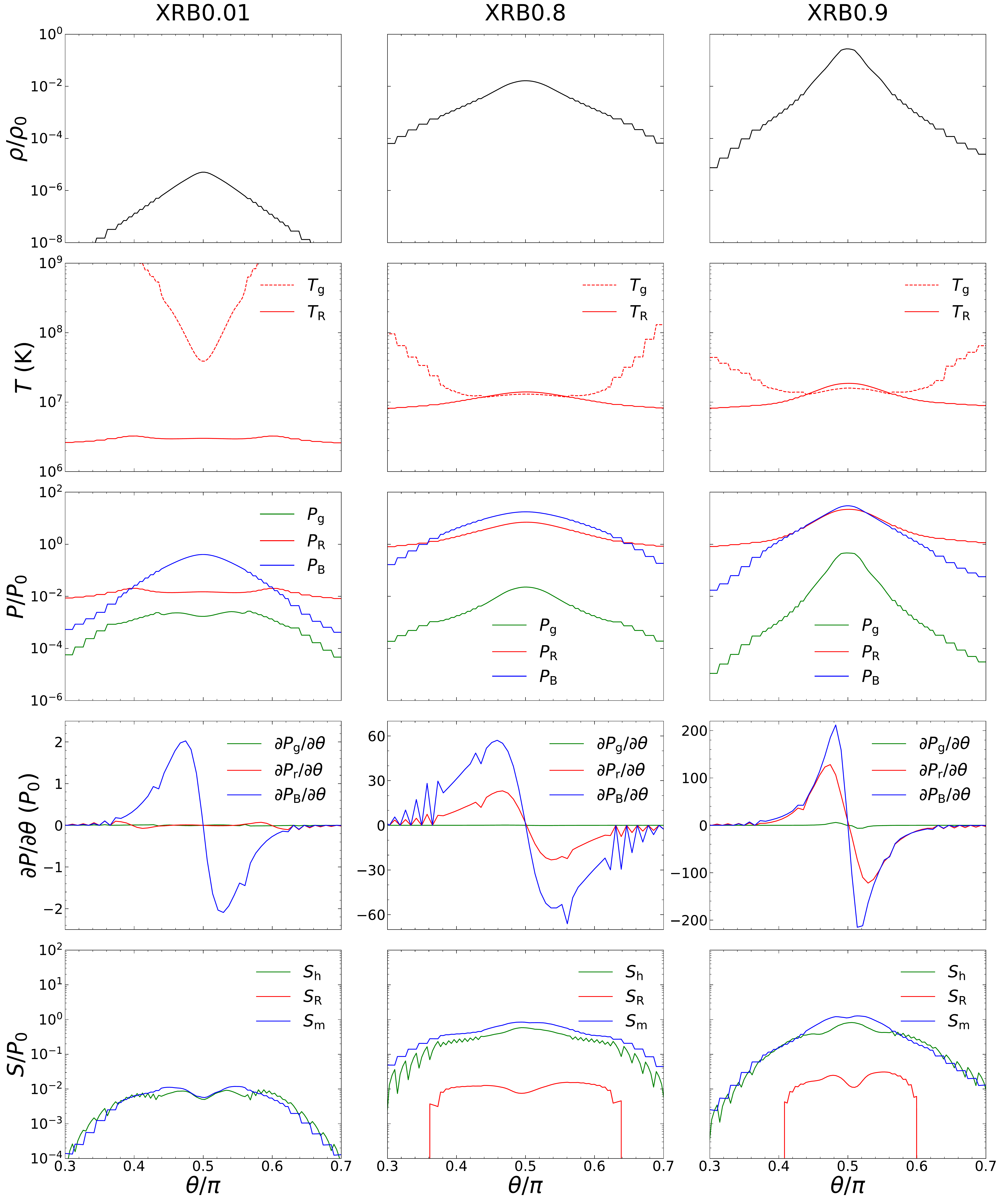}
\caption{Vertical profiles of density, temperatures, pressures, pressure gradients, and stresses at 10~$r_{\rm g}$ in the three runs. {\bf Top row}: vertical profiles of the gas density $\rho$. {\bf Second row}: vertical profiles of the gas temperature $T_{\rm g}$ (dashed red) and radiation temperature $T_{\rm R}$ (solid red). {\bf Third row}: vertical profiles of the gas pressure $P_{\rm g}$ (green), radiation pressure $P_{\rm R}$ (red), and magnetic pressure $P_{\rm B}$ (blue). {\bf Fourth row}: vertical profiles of the gas pressure gradient $\partial P_{\rm g}/\partial \theta$ (green), radiation pressure gradient $\partial P_{\rm R}/\partial \theta$ (red), and magnetic pressure gradient $\partial P_{\rm B}/\partial \theta$  (blue). {\bf Bottom row}: vertical profiles of the Reynolds stress $S_{\rm h}$ (green), radiation stress $S_{\rm R}$ (red), and Maxwell stress $S_{\rm m}$ (blue).
\label{fig:vertical_structure}}
\end{figure*}

Figure~\ref{fig:vertical_structure} shows the vertical (poloidal) structures of the accretion flow at 10~$r_{\rm g}$ in the three runs. The vertical density profiles of these three runs are similar, all peak at the disk mid plane and decrease towards the disk surface. The density decreases following an exponential relation $\rho \propto e^{-\left| z \right|}$, slower than a Gaussian profile $\rho\propto e^{-z^2}$ predicted for the isothermal case, and it is less concentrated on the disk mid-plane compared to the sub-Eddington accreting AGNs in \cite{Jiang2019}.

For the near-critical runs XRB0.8 and XRB0.9, the gas temperature and radiation temperature are in thermal equilibrium near the disk mid-plane. They peak at the mid-plane and decrease towards the disk surface. Above the disk surface, the radiation temperature continues to decrease, but the gas temperature starts to increase and is significantly higher than the radiation temperature in the funnel region. For the sub-critical run XRB0.01, the gas temperature is always higher than the radiation temperature, and the radiation temperature peaks near the disk surface. This is because the effective optical depth of the disk is low ($\textless$~10, see Figure \ref{fig:photon_trapping}). The radiation and gas have not reached local thermal equilibrium.

Although the radiation pressure is comparable to the magnetic pressure, the factor that supports the disk is actually the negative gradient of pressure from the disk mid-plane to the disk surface. In XRB0.01, the slope of the magnetic pressure is significantly higher than that of the radiation pressure; in XRB0.8, the former is higher by at least a factor of 2 than the latter; in XRB0.9, the former is higher by a factor of 2 than the latter near the disk mid-plane, but they become comparable at large angles. Thus, the disk is mainly supported by the magnetic pressure gradient near the disk mid-plane, which is similar to that found in the sub-Eddington simulation for an AGN \citep{Jiang2019}. The radiation pressure gradient becomes important when the accretion rate approaches the critical value, in particular at large scale heights.

\begin{figure}[ht]
\centering
\includegraphics[width=0.45\textwidth]{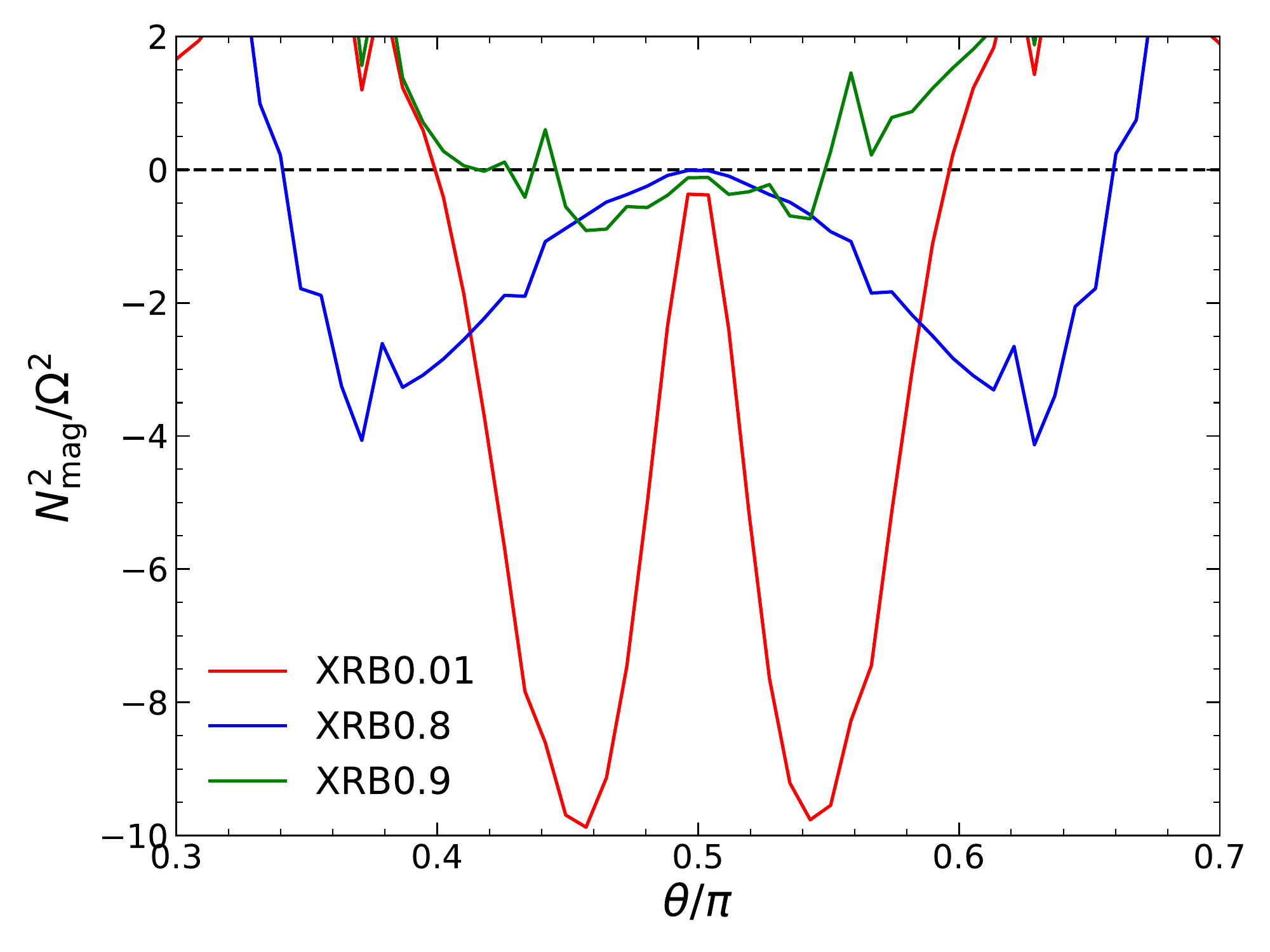}
\caption{Vertical profiles of the square of magnetic Brunt-V\"{a}is\"{a}l\"{a} frequency at 10~$r_{\rm g}$ for the three simulation runs.
\label{fig:Parker_mode}}
\end{figure}

\begin{figure}[ht]
\centering
\includegraphics[width=0.45\textwidth]{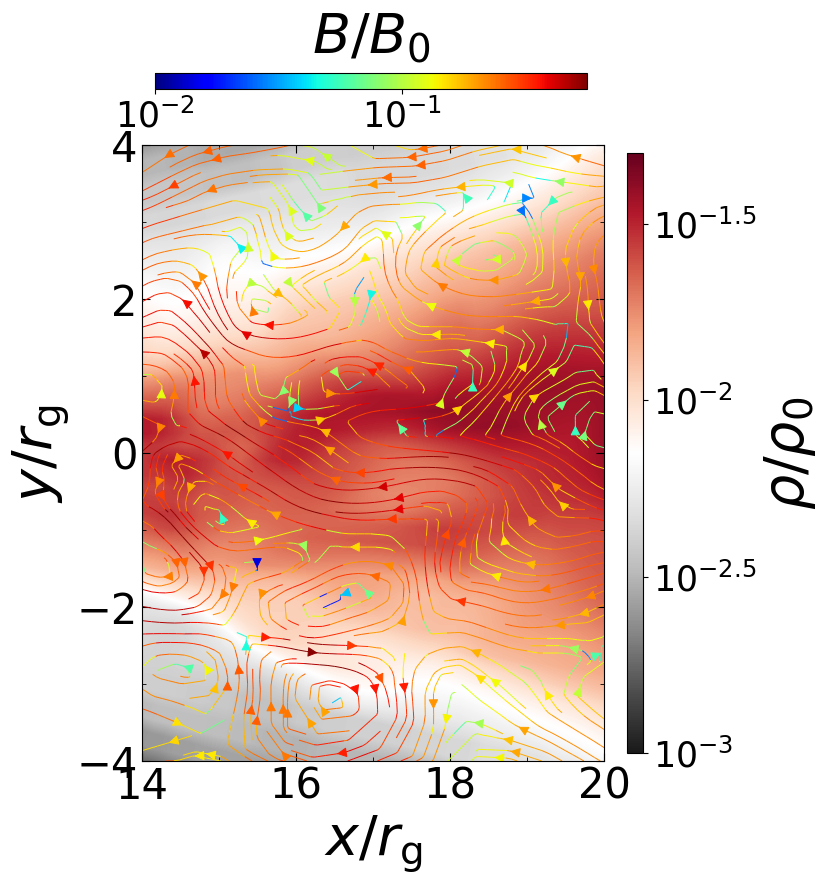}
\caption{Snapshot of density for the run XRB0.8 at the time 65~$t_0$ overlayed with magnetic field lines.
\label{fig:anti_correlation}}
\end{figure}

If the magnetic pressure gradient contributes significantly to supporting an accretion disk vertically, one may expect to see the undulatory Parker instability \citep{Tao2011}. To check that, we extract the square of the magnetic Brunt-V\"{a}is\"{a}l\"{a} frequency \citep{Blaes2011} expressed as
\begin{equation} 
N^2_{\rm mag} \equiv g \left( -\frac{g}{c^2_{\rm t}} - \frac{d \ln{\rho}}{dz} \right),
\end{equation}
where $g=\Omega^2\left| z \right|$ is the approximated acceleration due to the gravity of the central black hole, $c_{\rm t} \equiv \left[ \Gamma_1 \left( P_{\rm g} + P_{\rm R} \right) / \rho \right]^{1/2}$ is the adiabatic sound speed, and $\Gamma = 4/3$ is the adiabatic index. We show the vertical profiles of $N^2_{\rm mag} / \Omega^2$ for the three runs in Figure~\ref{fig:Parker_mode}. The values are negative at both sides near the mid-plane for the runs XRB0.01 and XRB0.8, suggesting the presence of undulatory Parker modes. We emphasize that the presence of Parker instability is not inconsistent with the disk being in a quasi-steady state, because the disk is not absolutely steady and the magnetic pressure is averaged over time. For the run XRB0.9, the values are negative only in a very limited vertical range, probably because the radiation pressure gradient has a non-negligible contribution in this run. For the run XRB0.8, we display a snapshot of density at the time 65~$t_{\rm 0}$ in Figure~\ref{fig:anti_correlation}. We zoom in the region near the mid-plane at radius 14--20~$r_{\rm g}$ to show the density fluctuations caused by Parker instability inside the disk. Ordered magnetic fields are apparent in low density regions while the turbulent component dominates the high density ones. The magnetic field strength, which is shown by the color of the stream lines, is anti-correlated with gas density. This is a clear signature of magnetic buoyancy \citep{Blaes2011, Jiang2014, Jiang2019a}.

The vertical profiles of the Maxwell and Reynolds stresses have a similar shape; both peak near the disk mid-plane and decline toward disk surfaces. The turbulent component of Maxwell stress has a small dip at the disk mid-plane in the run XRB0.01 because of the subtraction of the mean field component. The vertical profile of the radiation stress is different and shows a bimodal distribution. This is because the anisotropy of radiation determines the radiation stress, which reaches its maximum when the optical depth is close to unity. A similar bimodal distribution is also seen in sub-Eddington simulations around an AGN \citep{Jiang2019}, although the radiation stress in our case is not important.

\section{Discussion} 
\label{sec:discussion}

\subsection{Angular momentum distribution}
\label{subsec:angular_momentum}

\begin{figure}[ht]
\centering
\includegraphics[width=0.45\textwidth]{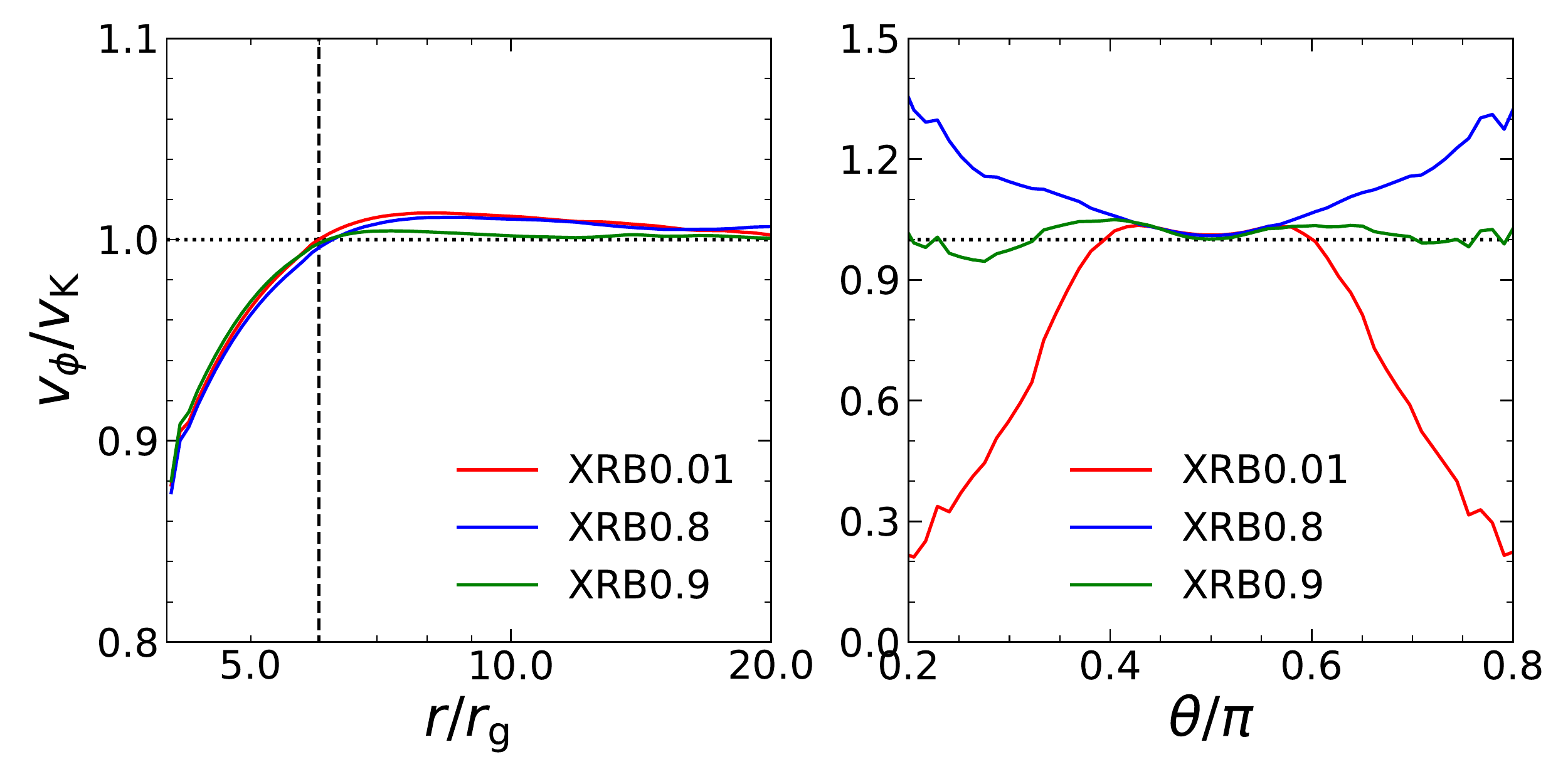}
\caption{Radial (left) and vertical (right) profiles of the density-weighted rotation velocity $v_{\phi}$ scaled with the Keplerian velocity $v_{\rm K}$. The black dashed line in the left panel marks the position of ISCO at 6~$r_{\rm g}$. 
\label{fig:angular_momentum}}
\end{figure}

The angular momentum distribution reflects how the matter is transferred into the central compact object. We plot the density-weighted rotation velocity as a function of radius for the three runs in Figure~\ref{fig:angular_momentum}. The disk motion is close to Keplerian outside the ISCO. These results are similar to those obtained with simulations in the cylindrical coordinates \citep{Jiang2014}. The radial rotation velocities are marginally super-Keplerian in all runs; this is because we have included the motion of outflows. We show the vertical profiles of rotation velocity at 10~$r_{\rm g}$ in the right panel of Figure~\ref{fig:angular_momentum}. Near the disk surface, where the outflow forms, the flow becomes super-Keplerian as the gravitational force cannot balance the centrifugal force. The gas motion is highly super-Keplerian for the run XRB0.8, and close to Keplerian for XRB0.9, consistent with the presence of outflows seen in Figure~\ref{fig:disk_structure}. The difference of XRB0.8 and XRB0.9 is caused by the different initial magnetic field but not the increasing mass accretion rate, because the trend is opposite to that from XRB0.01 to XRB0.8. In the run XRB0.01, there is no true outflows, and the motion above the scattering photosphere is dominated by the inflow of low density and low angular momentum initial density floor (see Figure~\ref{fig:disk_structure}). Below the scattering photosphere but above the effective absorption photosphere, the accretion gas dominates. This is the reason why the accretion flow is sub-Keplerian at high scale heights. The dominance of the Maxwell and Reynolds stresses over the radiation stresses in the three runs suggests that MRI plays an important role in angular momentum transfer \citep{Balbus1998}. 

\subsection{Comparison with theoretical disk models}
\label{subsec:disk_model}

 
Radiation produced inside the disk is released locally (Figure~\ref{fig:disk_structure}). The disks are nearly Keplerian in the bound gas region (Figure~\ref{fig:angular_momentum}). These satisfy the standard disk model assumptions. Contrary to the frequently-made assumption that it is constant throughout the disk, the ratio of vertically-integrated $r$--$\phi$ stress to vertically-integrated pressure (frequently called $\alpha$), drops almost an order of magnitude from the ISCO region, where it is $\sim$0.2, to 20~$r_{\rm g}$, where it is $\sim$0.03 (Figure~\ref{fig:radial_structure}). The scale height of a radiation pressure dominated $\alpha$ disk is determined by $H = \frac{\kappa_{\rm es} \dot{M}}{4 \pi c} \left| \frac{d\ln \Omega}{d\ln r} \right|$ in regions away from the innermost radius, as described in \citet{Shakura1973}. The disk scale height near the innermost radius is smaller by a factor $1-(r_{\rm in}/r)^{1/2}$. With $\dot{M}=0.0158\dot{M}_{\rm crit}$, $0.82\dot{M}_{\rm crit}$, and $0.9\dot{M}_{\rm crit}$ respectively for the runs XRB0.01, XRB0.8, and XRB0.9, the corresponding disk scale height predicted by the $\alpha$ disk model is $H=0.23r_{\rm g}$, $12r_{\rm g}$, and $14r_{\rm g}$ in the same order, independent of disk radius. However, the disks simulated in this work are thicker than the prediction of the $\alpha$ disk model. The scale height of effective absorption photosphere is proportional to the disk radius and it will exceed the model prediction at radius $\textgreater$10~$r_{\rm g}$. The thicker disk is consistent with that obtained from analytic analysis \citep{Begelman2007} and numerical simulation \citep{Sadowski2016a}. The standard disk model also ignores the magnetic field and assumes that the disk is gas or radiation pressure supported, and is predicted to be thermally unstable in the case when the radiation pressure dominates \citep{Shakura1976}. We find that these disks are actually magnetic pressure supported (Figure~\ref{fig:vertical_structure}). \citet{Begelman2007} assume a saturated magnetic pressure $P_{\rm B} \sim \rho c_{\rm s}v_{\rm K}$ according to the saturated Alfv\'{e}n velocity ($\sqrt{c_{\rm s}v_{\rm K}}$), where $c_{\rm s}$ is the gas sound speed and $v_{\rm K}$ is the Keplerian velocity. As one can see in Figure~\ref{fig:radial_structure}, their $P_{\rm B}$ roughly matches the simulated pressure in XRB0.01 and XRB0.9, and is lower than that in XRB0.8 by a factor of a few. The radiation efficiency is found to be $\sim$3--6\%, comparable to the prediction of standard disks.

Strong outflows with a velocity of $\sim$0.1$c$ are seen in runs XRB0.8 and XRB0.9.  Outflows are not included in both standard and slim disk models. However, radiation driven outflows are expected when the luminosity is high, especially when it approaches the Eddington limit \citep{Shakura1973,Watarai1999}.

\subsection{Disk properties as a function of accretion rate}

\begin{figure}[ht]
\centering
\includegraphics[width=\columnwidth]{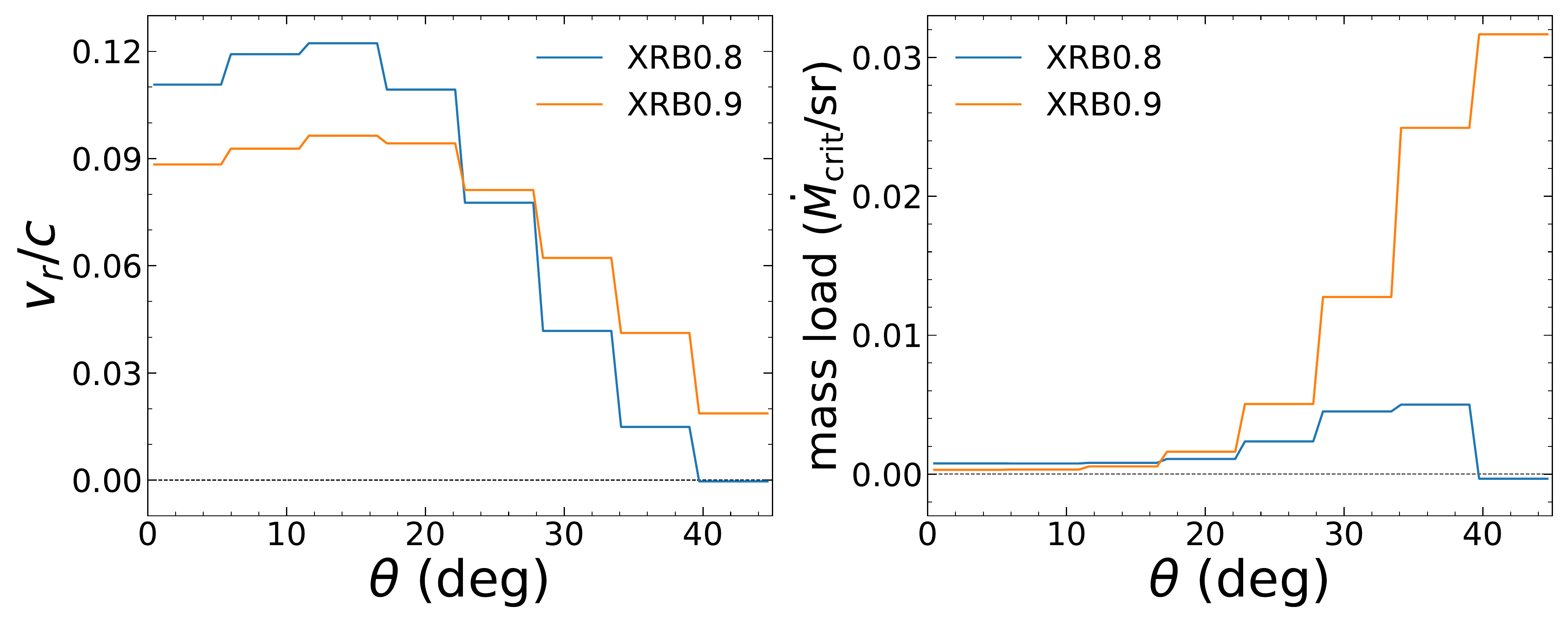}
\caption{Angular distribution of the outflow velocity and mass load in the runs XRB0.8 and XRB0.9. 
\label{fig:angular_distribution}}
\end{figure}

The three runs allow us to picture the evolution of accretion flow at different mass accretion rates.  We list some key properties of the disk in Table~\ref{tab:luminosity} as a function of $\dot{m}$.  The thickness of the disk ($\theta_{\rm d}$) is defined as the half opening angle of the effective absorption photosphere, which is also the half opening angle of the central low-density funnel.  The radiative temperature at mid-plane ($T_{\rm c}$) and on the effective photosphere ($T_{\rm ph}$) as a function of radius are fitted with a power-law function in the range of 10--20~$r_{\rm g}$ with a correction for zero torque at the innermost radius, i.e., $T(r) \propto r^p f^{1/4}$, where $f = 1 - (6r_{\rm g} / r) ^{1/2}$ and $p$ is the power-law index. We also show the angular distribution of outflow velocity and mass load, which is the mass loss rate per unit solid angle, in the funnel region for the two runs XRB0.8 and XRB0.9 in Figure~\ref{fig:angular_distribution}. The maximum wind velocity ($v_{\rm w}$) is summarized in Table~\ref{tab:luminosity} except for the run XRB0.01. These relations may help develop a semi-analytic accretion disk model that takes into account both advection and outflows.

As the accretion rate increases, the accretion disk becomes thicker at a given radius, and the radial profile of radiation temperature becomes flatter both at the mid-plane and on the effective absorption photosphere. The mid-plane temperature profile in the run XRB0.9 seems not to follow such a trend but becomes steeper than XRB0.8, because it cannot be well described by the radial model. The temperature profiles are all flatter than that predicted by the standard accretion disk ($p = -0.75$),  but close to that predicted by slim disk ($p = -0.5$) except for $T_{\rm c}$ in XRB0.8 and $T_{\rm ph}$ in XRB0.9, which may be a result of strong advection in the accretion flow. We adopt the gas temperature at 10~$r_{\rm g}$ on the rotational axis as an estimation of the corona temperature ($T_{\rm axis}$). The corona cools with increasing accretion rate, with a temperature of $3\times10^9$~K, $2\times10^9$~K, and $8\times10^8$~K, respectively, in XRB0.01, XRB0.8, and XRB0.9. The radiation efficiency depends weakly upon the accretion rate. It decreases by a factor of 2 when the accretion rate increases from 1\% to 80-90\% of the critical value. Also, as the accretion rate increases, more outflows are launched, and thus the outflow to net inflow rate increases with increasing accretion rate. The outflow velocity is $\sim$0.1$c$ and does not seem to vary with accretion rate. For the two near-critical runs (XRB0.8 and XRB0.9), the outflow velocity decreases with increasing inclination angle ($\theta$), while the mass load increases with it. As a result, an observer along the funnel edge sees more outflows than along the rotational axis.

%

\subsection{Outflows} 

In Table~\ref{tab:luminosity}, we show that in the near-critical run XRB0.8 and XRB0.9, the rate for true outflows that will escape to infinity is $\sim$0.02~$\dot{M}_{\rm crit}$. The ratio of true outflow rate to net mass accretion rate is $\sim$3\%. In the large-scale RHD simulations for supercritical accretion \citep{Kitaki2021}, they find that the true outflow rate is 2.4~$\dot{M}_{\rm crit}$ given a supercritical net mass accretion rate of 18~$\dot{M}_{\rm crit}$ (converted according to our definition). Their ratio of true outflow rate to net mass accretion rate is $\sim$13\%, which is consistent with the trend we show in Table \ref{tab:luminosity} that the ratio will grow with the increasing net accretion rate.

\section{Conclusions}
\label{sec:conclusion}

In this paper, we present results with 3D global RMHD simulations of accretion onto a 6.62~$M_\sun$ black hole, with different initial magnetic configurations and consequently different accretion rates from a few percent to $\sim\dot{M}_{\rm crit}$. Main results are summarized below.

Outflows start from the ISCO and the mass loss rate increases rapidly with radius. We see no outflow when the accretion rate is about $10^{-2}$~$\dot{M}_{\rm crit}$. The true outflow to net accretion rate is around 2.5\% when the net accretion rate reaches near the critical rate. The ratio of true outflow rate to the net mass accretion rate increases with the mass accretion rate. The peak velocity of the outflow is about 0.1~$c$ and the mass load of the outflow is peaked near the disk surface.

In the near-critical accretion flow, the energy dissipation occurs mainly inside the disk, while in the sub-critical case, the energy dissipates mainly at the disk surface. The radial velocity is over 10 times the vertical photon diffusion speed in the runs XRB0.8 and XRB0.9, meaning photon trapping is important in the near-critical accretion flow. The ratio of radial velocity to the vertical photon diffusion speed increases with increasing accretion rate, indicating the photon trapping effect is more obvious when the mass accretion rate is higher. The radiation efficiency is a few percent and depends weakly upon the accretion rate. It decreases by a factor of 2 when the accretion rate increases from 1\% to 80-90\% of the critical value.

The disk is dominated by the magnetic pressure. The magnetic pressure is larger or comparable to the radiation pressure. The negative vertical gradient of magnetic pressure is significantly larger or at least larger by a factor of 2 than the radiation pressure within a few scale heights from the mid-plane, indicating the disk is magnetic pressure supported vertically. The value of magnetic pressure can be roughly described by the saturated magnetic pressure as presented in \citet{Begelman2007}. The Maxwell and Reynolds stresses are the main sources of angular momentum transfer, which is in contrast with the AGN case, where the radiation stress plays an important role \cite{Jiang2019}.

\section*{Acknowledgements}
We thank the anonymous referee for useful comments that help improve the manuscript. HF acknowledges funding support from the National Key R\&D Project under grant 2018YFA0404502, the National Natural Science Foundation of China under grants Nos.\ 12025301 \& 11821303, and the Tsinghua University Initiative Scientific Research Program. An award of computer time was provided by the Innovative and Novel Computational Impact on Theory and Experiment (INCITE) program. This research used resources of the Argonne Leadership Computing Facility, which is a DOE Office of Science User Facility supported under Contract DE-AC02-06CH11357. Part of this work was performed using resources provided by the Cambridge Service for Data Driven Discovery (CSD3) operated by the University of Cambridge Research Computing Service (www.csd3.cam.ac.uk), provided by Dell EMC and Intel using Tier-2 funding from the Engineering and Physical Sciences Research Council (capital grant EP/T022159/1), and DiRAC funding from the Science and Technology Facilities Council (www.dirac.ac.uk). The Center for Computational Astrophysics at the Flatiron Institute is supported by the Simons Foundation. JS acknowledges support from a NASA TCAN grant 80NSSC21K0496. MM acknowledges support via an STFC Consolidated grant (ST/V001000/1). SWD acknowledges support from NASA Astrophysics Theory Program grant 80NSSC18K1018.



\end{CJK*}
\end{document}